\documentclass[twocolumn,numberedappendix,appendixfloats]{emulateapj}

\usepackage{amsmath}
\usepackage{amssymb}
\usepackage{graphicx}
\usepackage{natbib}
\newcommand{\Iwe}{we}
\newcommand{\IWe}{We}
\newcommand{\myour}{our}
\newcommand{\MyOur}{Our}

\newcommand{\dnu}{\ensuremath{\,\mathrm{d}\nu}}

\newcommand{\dex}[2]{\ensuremath{#1 \cdot 10^{#2}}}
\newcommand{\Teff}{\ensuremath{T_\mathrm{eff}}}
\newcommand{\Kelvin}{\ensuremath{\,\mathrm{K}}}
\newcommand{\Msun}{\ensuremath{M_{\sun}}}
\newcommand{\Rsun}{\ensuremath{R_{\sun}}}
\newcommand{\Lsun}{\ensuremath{L_{\sun}}}
\newcommand{\fotsek}{\ensuremath{\text{photons}\cdot{\text s}^{-1}}}
\newcommand{\TLUSTY}{{\tt TLUSTY}}
\newcommand{\CMFGEN}{{\tt CMFGEN}}
\newcommand{\PHOENIX}{{\tt PHOENIX}}
\newcommand{\FASTWIND}{{\tt FASTWIND}}
\newcommand{\WMBASIC}{{\tt WMBasic}}
\newcommand{\TMAP}{{\tt TMAP}}

\DeclareMathAlphabet{\mathsc}{OT1}{cmr}{m}{sc}
\def\testbx{bx}%
\DeclareRobustCommand{\ion}[2]{%
\relax\ifmmode
\ifx\testbx\f@series
{\mathbf{#1\,\mathsc{#2}}}\else
{\mathrm{#1\,\mathsc{#2}}}\fi
\else{#1\,{\scshape{#2}}}%
\fi}

\begin{document}
\title{Spherically symmetric NLTE model atmospheres of hot
hydrogen-helium first stars}
\shorttitle{NLTE model atmospheres of first stars}
\author{Ji\v{r}\'{\i} Kub\'{a}t}
\shortauthors{Kub\'at}

\affil{Astronomick\'y \'ustav,
	Akademie v\v{e}d \v{C}esk\'e republiky,
	CZ-251 65 Ond\v{r}ejov, Czech Republic
	}

\begin{abstract}
{\IWe} present results of {\myour} calculations of NLTE model stellar
atmospheres for hot Population III stars composed of hydrogen and
helium.
{\IWe} use {\myour} own computer code for calculation of spherically
symmetric NLTE model atmospheres in hydrostatic and radiative
equilibrium.
The model atmospheres are then used for calculation of emergent fluxes.
These fluxes serve for the evaluation of the flow of high-energy photons
for energies higher than ionization energies of hydrogen and helium, the
so-called ionizing photon fluxes.
{\IWe} also present the time evolution of the ionizing photon fluxes.
\end{abstract}


\section{Introduction}

At the beginning of the era of modern NLTE stellar atmosphere modelling
at the early 70's, the chemical composition of computed model
atmospheres had to be as simple as possible to enable their calculation
at computers that were available at that time.
Therefore, these models included only minimum number of elements,
selected according to their importance for the atmospheric structure
properties.
First models were purely hydrogen \citep{amnlte3,amnlte4}, later also
helium was included
\citep{amnlte5,amnlte7,1973A&A....28..103K,1976A&A....52...11K}
to calculations.
Due to computer limitations, the absorption by heavier elements was not
considered, and, of course, also the absorption by a huge number of
metal lines in the ultraviolet spectral region was omitted.
These metal lines cause lowering the ultraviolet flux and enhancing the
flux in the visual region.
This important effect is known as line blanketing \citep[see,
e.g.,][]{sa2}.
Great effort \citep[started by][]{mfmg1,mfmg2} was necessary to include
the line blanketing into NLTE model atmospheres calculations, and
nowadays it is possible to calculate NLTE model atmospheres with this
effect taken into account properly.
One of the most widely used codes for calculation of static NLTE line
blanketed plane-parallel model atmospheres is the code {\TLUSTY}
\citep{tlustypopis,nlteblan1}.
NLTE line-blanketing can be also treated by other model atmosphere
codes, like the model atmosphere package
{\TMAP} \citep{2003ASPC..288...31W}.
For the case of NLTE moving atmospheres with line blanketing, codes
{\CMFGEN} \citep{cmfgen}, {\PHOENIX} \citep[e.g.][]{phoenixpar},
{\FASTWIND} \citep{fastwind1,fastwind2},
and {\WMBASIC} \citep[e.g.][and references therein]{2003RvMA...16..133P}
are available.

All these mentioned model atmosphere codes were developed and used with
the primary objective to reproduce the observed emergent radiation from
hot stars.
The hydrogen-helium static model atmospheres calculated by
\citeauthor{amnlte3} served only as a first approximation of NLTE hot
stellar atmosphere modelling.
However, there is a group of stars, whose chemical composition
corresponds exactly to what was calculated by \citeauthor{amnlte3} in
the early 70's, namely the first stars in the Universe.
Since the Universe at its beginning was practically free of metals and
consisted only of hydrogen, helium, and a very little amount of lithium
\citep[see, e.g., reviews of][]
{2004ARA&A..42...79B,2008IAUS..250..471J,2009Natur.459...49B},
the first stars may be safely assumed to consist only of hydrogen and
helium, since the influence of lithium on the stellar atmospheric
structure is negligible.

Thus, in order to obtain model atmospheres of first stars, it is simply
sufficient to use the first codes for calculation of NLTE model
atmospheres from the early 70's, or to repeat such calculations with
contemporary sophisticated model atmosphere codes and just to skip
metals from the calculations.
The latter was done by \cite{2000ApJ...528L..65T}, who used the code
{\TLUSTY} to calculate simple hydrogen-helium model atmospheres and
rediscovered that {\em ``A model atmosphere is necessary because a
simple blackbody curve for each {\Teff} will not accurately reproduce
the spectrum ... ''} \citep[cf.][where the history of the finding that
the stellar radiation is not the blackbody one is thoroughly
descibed]{unsold}.
They concluded that neglecting all the metal opacities leads to
enhancement of the UV flux emerging from the model stellar atmosphere, a
fact that was well known before, but in opposite formulation, namely
that adding metallic opacity reduces the ultraviolet stellar flux.

The presence of high UV flux emerging from the hydrogen-helium model
atmospheres
describes an important source of radiation, which is able to ionize the
intergalactic medium in the early Universe evolution
\citep{2003ApJ...584..621V}.
Indeed, high fluxes at high frequencies caused by missing stellar
opacity in those spectral regions are extremely important in cosmology
\citep{2003ApJ...594L...1V}.
Model atmospheres of first stars and their influence on ionizing
radiation were also studied in detail by \cite{ds02,ds03}.

The absence of metals in the atmospheres of first stars has an
additional effect.
Since the winds of hot stars are driven by radiation scattered by
spectral lines, there was a question whether the first stars have
stellar winds.
Existence or non-existence of the stellar wind has dramatic influence on
the first star evolution.
\cite*{2001ApJ...552..464B} calculated unified model atmospheres
\citep[using an improved version of the method of][]{fastwind1} and did
not find any effect of the velocity field on line profiles for low
mass-loss rates.
\cite{2002ApJ...577..389K} tested the existence of winds of first stars
by calculating wind models using the depth dependent line force
multiplier parameters $k$, $\alpha$, and $\delta$.
He tried to mimic the zero metallicity of the first stars using a
method, which implicitly assumes driving of the wind by metalic lines.
He pointed out the possibility of multicomponent effects in the wind of
first stars.
These were later studied in detail by \cite{prvnih,petcno}.
However, the principal question of existence or non-existence of the
winds of first stars deserved more detailed study.
The code of \cite{nlte1}
enabled calculation of the radiative force directly using actual level
populations and opacities of individual elements at each depth point,
without the necessity to introduce parameterization of the radiative
force by means of the line force multipliers.
Consequently, this method gives implicitly depth-dependent line force
and allows to include consistently any chemical composition.
Using this code, \cite{bezvi,cnovit} studied stellar winds of first
stars and
confirmed the fact found by \cite{2002ApJ...577..389K}
that the first stars
have extremely weak winds.
These calculations also showed that using static approximation for
calculation of model atmospheres of first stars is adequate.
For these calculations, which were done using the core-halo
approximation, it was necessary to determine the flux at the lower
boundary of the wind region.
To this end, the emergent radiation from the static spherically
symmetric NLTE model atmospheres was used.
Consequently, \citeauthor{cnovit} did not check if the plane-parallel
approximation, which has been frequently used by others, is adequate for
modelling the atmospheres of first stars.
Since the NLTE atmospheric models used by \citeauthor{cnovit} have not
been published anywhere in detail,
in this paper {\Iwe} present an extended set of NLTE model atmospheres used
for this purpose.

\section{Model atmospheres}

\subsection{Validity of static approximation}

For the case when the thickness of the atmosphere is small compared to
the stellar radius, the plane-parallel approximation offers a good
choice.
This happens for dwarf stars with relatively high surface gravitational
accceleration ($\log g \gtrsim 4$).
However, small deviations from the plane-parallel approximation were
found even for subdwarfs \citep{gruschinske78}.
For the case of white dwarfs they were found only in cores of some
strong lines and were usually very small \citep{spwd}.

On the other hand, for Population I and II stars with lower surface
gravities, giants and supergiants, the atmosphere is more extended, and
it can no longer be considered as thin with respect to the stellar
radius.
In such case a spherically symmetric approximation is more realistic
than the plane-parallel one.
However, due to a weaker gravitational force in the atmospheres of
giants and supergiants, the radiation force may easily overcome gravity
there and give rise to a stellar wind.
Consequently, such atmosphere can not be considered as a static one.
Winds in hot star atmospheres are driven by radiation force generated by
momentum transfer from absorptions and scatterings in lines of heavy
elements, with a help of the force generated by continuum transitions
and electron scattering.
Note that the force generated by continuum transitions and electron
scattering alone is not sufficient to drive the wind.

The first stars in Universe were born from pure hydrogen-helium mixture
with no heavier elements.
Consequently, since the radiation force does not overcome gravity there,
the line driven wind does not exist there \cite[see][]{bezvi,cnovit} and
atmospheres of these stars may be described using static approximation
in both plane-parallel and spherically symmetric geometry.

\subsection{Method of calculation}

{\MyOur} method for calculation of NLTE model stellar atmospheres is
based on an accelerated lambda iteration (ALI) method and has been
described in some detail in \cite{ATA1,ATA2,ATA3,ATA4} and summarized in
\cite{ATAsum}.
Using {\myour} code {\Iwe} calculated a grid of hydrogen-helium
spherically symmetric NLTE model atmospheres in hydrostatic and
radiative equilibrium.
The helium abundance $Y_\mathrm{He}=n_\mathrm{He}/n_\mathrm{H}=0.1$.
{\IWe} considered a 16 level hydrogen atom (15 levels of \ion{H}{i} + 1
level of \ion{H}{ii}) and a 50 level helium model atom (29 levels of
\ion{He}{i} + 20 levels of \ion{He}{ii} + 1 level of \ion{He}{iii}).
Individual hydrogen levels correspond to main quantum numbers
$n=1,\dots,15$.
For \ion{He}{i}, all levels up to $n=4$ were taken separately according
to orbital quantum number $l$, for $5\le n \le 9$ {\Iwe} took 2 averaged
levels for each $n$, one for singlets, one for triplets.
The levels of \ion{He}{ii} were considered similarly as the hydrogen
levels, however for $n=1,\dots,20$.
Details of the model atoms used (oscillator strengths, photoionization
cross sections, collisional excitation and ionization rates, line
profiles) can be found in \cite{tt-kpp}.

For temperature structure determination we used a combination of three
methods.
At large depths below the stellar radius $R_*=r(\tau_R\approx
2/3)$ ($\tau_R$ is the Rosseland optical depth) it was the differential
form of the radiative equilibrium equation, above the stellar radius the
integral form of radiative equilibrium equation, and for outermost
layers the electron thermal balance method was used
\citep[for details see][]{ATA2,ATA4,tt-kpp}.

\section{Results of calculations}

\subsection{Model atmosphere parameters}

Basic parameters (luminosity $L_*$, mass $M_*$, and radius $R_*$) of the
first stars are taken from evolutionary calculations.
For the original calculations used in \cite{cnovit} a limited set
of model parameters \citep[taken from][]{zermetevol} was used.
In addition to them, here we extended the set by using additional
parameters from \citeauthor{zermetevol}, and we also used other sources
of stellar parameters following from other evolutionary calculations.
Values of masses of these stars are still matter of debate and they are
different in different calculations
(cf. different results obtained by evolutionary calculations of
\citealt{hol1} or \citealt{hol2}, and \citealt{verylowmet}).

\begin{table*}
\centering
\caption{Model atmosphere parameters and and stellar ionizing photon
fluxes $Q_i$ for selected model atmospheres of first stars with
parameters from \cite{ds02}.}
\label{qqds}
\begin{tabular}{llrlrl|lll}
\hline
model & $R_*$ & $M_*$ & $L$ &
\Teff & $\log g$ 
& $Q(\ion{H}{i})$
& $Q(\ion{He}{i})$
& $Q(\ion{He}{ii})$
\\
& [\Rsun] & [\Msun] & [\Lsun] & [\Kelvin] & &
\multicolumn{3}{c}{[\fotsek]} \\
\hline
ds999 & 15.57 & 1000. & \dex{2.780}{7} & 106170 & 5.053
& \dex{1.25}{50} & \dex{8.59}{49} & \dex{4.81}{49}
\\
ds500 & 10.41 & 500. & \dex{1.276}{7} & 106900 & 5.102
& \dex{5.66}{49} & \dex{3.85}{49} & \dex{2.08}{49}
\\
ds400 & 9.09 & 400. & \dex{9.638}{6} & 106660 & 5.123
& \dex{4.39}{49} & \dex{2.99}{49} & \dex{1.61}{49}
\\
ds300 & 8.28 & 300. & \dex{6.592}{6} & 101620 & 5.079
& \dex{3.12}{49} & \dex{2.06}{49} & \dex{1.04}{49}
\\
ds200 & 6.48 & 200. & \dex{3.750}{6} & 99770 & 5.116
& \dex{1.84}{49} & \dex{1.20}{49} & \dex{5.75}{48}
\\
ds120 & 4.81 & 120. & \dex{1.750}{6} & 95720 & 5.153
& \dex{9.02}{48} & \dex{5.67}{48} & \dex{2.47}{48}
\\
ds080 & 3.60 &  80. & \dex{8.851}{5} & 93320 & 5.230
& \dex{4.71}{48} & \dex{2.90}{48} & \dex{1.17}{48}
\\
ds060 & 3.12 &  60. & \dex{5.188}{5} & 87700 & 5.228
& \dex{2.87}{48} & \dex{1.69}{48} & \dex{6.03}{47}
\\
ds040 & 2.71 &  40. & \dex{2.630}{5} & 79430 & 5.175
& \dex{1.48}{48} & \dex{8.22}{47} & \dex{2.56}{47}
\\
ds025 & 1.85 &  25. & \dex{7.762}{4} & 70800 & 5.301
& \dex{4.27}{47} & \dex{2.38}{47} & \dex{6.44}{46}
\\
ds015 & 1.47 &  15. & \dex{2.109}{4} & 57410 & 5.281
& \dex{1.07}{47} & \dex{5.19}{46} & \dex{1.17}{46}
\\
ds009 & 1.36 &   9. & \dex{5.117}{3} & 41880 & 5.127
& \dex{1.79}{46} & \dex{5.69}{45} & \dex{9.95}{44}
\\
ds005 & 1.20 &   5. & \dex{7.413}{2} & 27540 & 4.982
& \dex{6.07}{43} & \dex{3.28}{40} & \dex{1.65}{37}
\\
\hline
\end{tabular}
\end{table*}

\begin{figure}
\resizebox{\hsize}{!}{\includegraphics{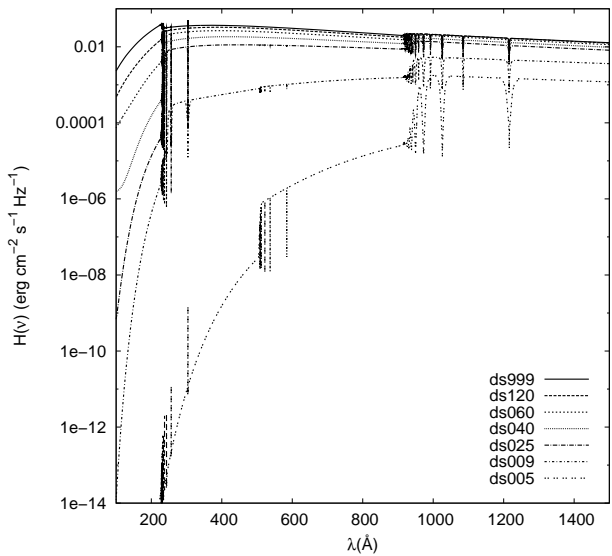}}
\caption{Plot of ultraviolet fluxes from selected model atmospheres of
first stars with parameters from \cite{ds02}.
Labels correspond to the first column of the Table~\ref{qqds}.}
\label{dsflux}
\end{figure}

First stars model atmospheres calculated by \cite{ds02} cover stellar
masses from 5\,{\Msun} up to extremely high masses of 1000\,\Msun.
He used the code {\TLUSTY} for calculation of H-He model atmospheres for
stars with $\Teff>20000{\Kelvin}$, for colder stars he adopted the
Kurucz LTE model atmospheres with a very low metallicity.
{\IWe} calculated spherically symmetric NLTE model atmospheres for
parameters given in the Table~3 of \cite{ds02},
these parameters are listed here in the Table~\ref{qqds}, corresponding
temperature structures
as a function of the column mass depth $m$
are shown in the Figure~\ref{dstemp}.
The column mass depth is related to the Rosseland optical depth by a
relation
\begin{equation}\label{colmade}
\text{d}m = (R_*^2/r^2) (\rho/\chi_\text{R}) \text{d}\tau_\text{R},
\end{equation}
where $\rho$ is the density and $\chi_\text{R}$ the Rosseland opacity.
All models display a rise of temperature in the outer parts of the
atmosphere, where the strongest lines are optically thick and the
continuum is optically thin.
This is a typical NLTE effect in model atmospheres without metals
\citep[see, e.g.][]{amnlte4}.
In model atmospheres with metal line blanketing included, this
temperature rise disappears \citep[e.g.,][]{BSTAR}.

To check the validity of the plane-parallel approximation for the first
stars by direct comparison of spherically symmetric and plane-parallel
models calculated with the same code (which has not been done yet),
{\Iwe} also calculated NLTE plane-parallel model atmospheres for these
stellar parameters.
{\IWe} found that the difference between the temperature structures of
spherically symmetric and plane-parallel model atmospheres is relatively
small, similarly to other cases where the atmospheric extension is not
too large \cite[cf.][] {gruschinske78,ATA2,scspn,sphds,spagb}.
Also the difference between emergent fluxes from spherically symmetric
and plane-parallel model atmospheres (the sphericity effect) is small,
the flux from the spherically symmetric model atmosphere is slightly
lower.

Although it was initially assumed that Population III stars should have
very large masses of the order of hundreds \Msun,
\cite{zermetevol}
calculated an extensive grid of evolutionary models of
Population III stars with masses $0.7\,M_{\sun}\le M\le 100\,M_{\sun}$
with the assumption of zero mass-loss.

\begin{table*}
\centering
\caption{Model atmosphere parameters and and stellar ionizing photon
fluxes $Q_i$ for selected model atmospheres for parameters from
\cite{zermetevol}.}
\label{qqzams}
\begin{tabular}{llrlll|lll}
\hline
model & $R_*$ & $M_*$ & $L$ &
\Teff & $\log g$ 
& $Q(\ion{H}{i})$
& $Q(\ion{He}{i})$
& $Q(\ion{He}{ii})$
\\
& [\Rsun] & [\Msun] & [\Lsun] & [\Kelvin] & &
\multicolumn{3}{c}{[\fotsek]}\\
\hline
M999 & 4.23 & 100 & \dex{1.282}{6} & 94400 & 5.185
& \dex{6.74}{48} & \dex{4.18}{48} & \dex{1.74}{48}
\\
M700 & 3.44 &  70 & \dex{6.871}{5} & 89530 & 5.209
& \dex{3.72}{48} & \dex{2.26}{48} & \dex{8.31}{47}
\\
M500 & 2.82 &  50 & \dex{3.597}{5} & 84140 & 5.236
& \dex{2.01}{48} & \dex{1.16}{48} & \dex{3.80}{47}
\\
M300 & 2.10 &  30 & \dex{1.189}{5} & 73960 & 5.271
& \dex{6.61}{47} & \dex{3.77}{47} & \dex{1.05}{47} 
\\
M200 & 1.65 &  20 & \dex{4.456}{4} & 65310 & 5.305
& \dex{2.40}{47} & \dex{1.27}{47} & \dex{3.24}{46}
\\
M150 & 1.48 &  15 & \dex{2.128}{4} & 57280 & 5.273
& \dex{1.08}{47} & \dex{5.18}{46} & \dex{1.16}{46}
\\
M120 & 1.42 &  12 & \dex{1.132}{4} & 49890 & 5.210
& \dex{5.00}{46} & \dex{2.10}{46} & \dex{4.04}{45}
\\
M100 & 1.37 &  10 & \dex{6.653}{3} & 44460 & 5.162
& \dex{2.48}{46} & \dex{8.36}{45} & \dex{1.39}{45}
\\
M095 & 1.36 & 9.5 & \dex{5.727}{3} & 43050 & 5.15
& \dex{1.99}{46} & \dex{5.94}{45} & \dex{9.92}{44}
\\
M090 & 1.34 &   9 & \dex{4.875}{3} & 41590 & 5.135
& \dex{1.55}{46} & \dex{4.13}{45} & \dex{6.71}{44}
\\
M083 & 1.32 & 8.3 & \dex{3.810}{3} & 39440 & 5.11
& \dex{1.03}{46} & \dex{1.93}{45} & \dex{2.49}{44}
\\
M080 & 1.31 &   8 & \dex{3.412}{3} & 38460 & 5.103
& \dex{8.34}{45} & \dex{1.30}{45} & \dex{1.60}{44}
\\
M070 & 1.30 &   7 & \dex{2.234}{3} & 34830 & 5.057
& \dex{2.90}{45} & \dex{5.68}{43} & \dex{2.14}{42}
\\
M060 & 1.27 &   6 & \dex{1.426}{3} & 31400 & 5.005
& \dex{5.73}{44} & \dex{1.02}{42} & \dex{8.98}{38}
\\
M050 & 1.23 &   5 & \dex{8.054}{2} & 27670 & 4.954
& \dex{7.23}{43} & \dex{4.01}{40} & \dex{1.93}{37}
\\
M040 & 1.17 &   4 & \dex{3.837}{2} & 23600 & 4.903
& \dex{8.65}{42} & \dex{1.82}{39} & \dex{1.47}{35}
\\
M030 & 1.11 &   3 & \dex{1.476}{2} & 19050 & 4.821 
& \dex{5.71}{41} & \dex{1.39}{37} & \dex{4.75}{31}
\\
M022 & 1.02 & 2.2 & \dex{4.497}{1} & 14820 & 4.77
& \dex{2.07}{40} & \dex{8.08}{34} & \dex{1.44}{28}
\\
M021 & 1.01 & 2.1 & \dex{3.767}{1} & 14220 & 4.75
& \dex{9.83}{39} & \dex{3.38}{34} & \dex{3.16}{27}
\\
M020 & 0.996&   2 & \dex{3.133}{1} & 13680 & 4.742
& \dex{6.69}{39} & \dex{2.71}{34} & \dex{1.68}{27}
\\ 
M019 & 0.982 & 1.9 & \dex{2.552}{1} & 13090 & 4.73 
& \dex{3.67}{39} & \dex{1.91}{34} & \dex{8.43}{26}
\\
M018 & 0.980 & 1.8 & \dex{2.080}{1} & 12440 & 4.71
& \dex{5.19}{39} & \dex{3.89}{34} & \dex{1.93}{27}
\\
M017 & 0.959 & 1.7 & \dex{1.641}{1} & 11860 & 4.70
& \dex{1.18}{39} & \dex{1.03}{34} & \dex{1.12}{26}
\\
M016 & 0.944 & 1.6 & \dex{1.274}{1} & 11220 & 4.69
& \dex{3.75}{38} & \dex{4.10}{32} & \dex{1.49}{24}
\\
M015 & 0.932 & 1.5 & \dex{9.772}{0} & 10570 & 4.68
& \dex{3.66}{38} & \dex{4.00}{32} & \dex{1.45}{24}
\\
\hline
\end{tabular}
\end{table*}

\begin{figure}
\resizebox{\hsize}{!}{\includegraphics{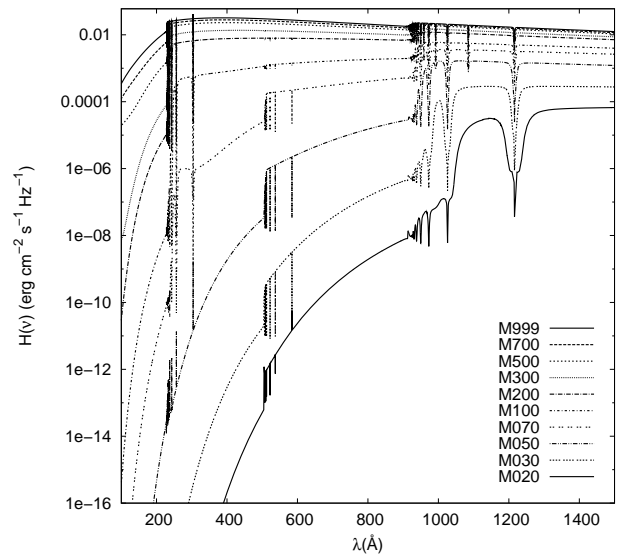}}
\caption{Plot of ultraviolet fluxes from selected model atmospheres of
zero-age main sequence first stars from \citet{zermetevol}. Labels
correspond to the first column of the Table~\ref{qqzams}.}
\label{zamsflux}
\end{figure}

For selected masses from this grid {\Iwe} calculated NLTE model
atmospheres for models of stars with zero age.
Their parameters ($L_*$, $M_*$, $R_*$) were taken from the online Table
in \cite{zermetevol} and are listed in the Table~\ref{qqzams}.
Temperature structures for selected models are plotted in the
Figure~\ref{zamstemp}.
All models display the same type of the rise of temperature as the
models in the Figure~\ref{dstemp}.

\subsection{Ionizing fluxes and their evolution}

\begin{figure}
\resizebox{\hsize}{!}{\includegraphics{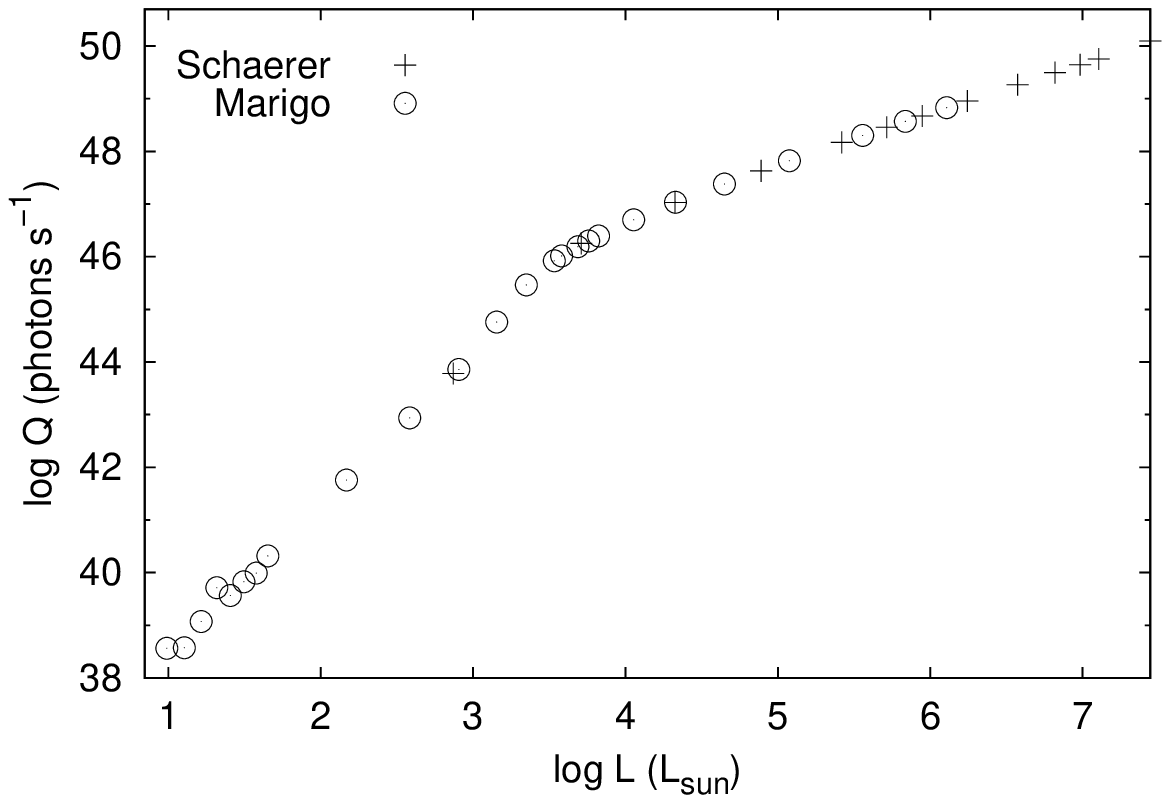}}
\resizebox{\hsize}{!}{\includegraphics{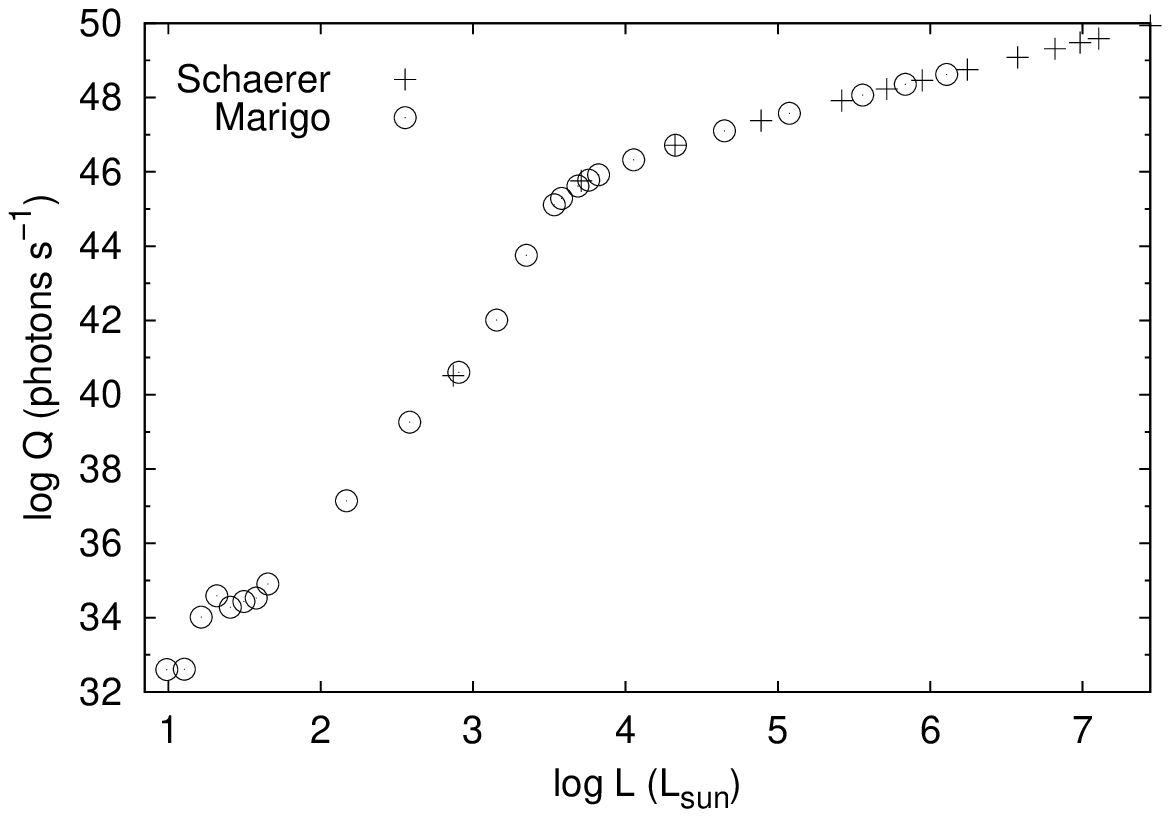}}
\resizebox{\hsize}{!}{\includegraphics{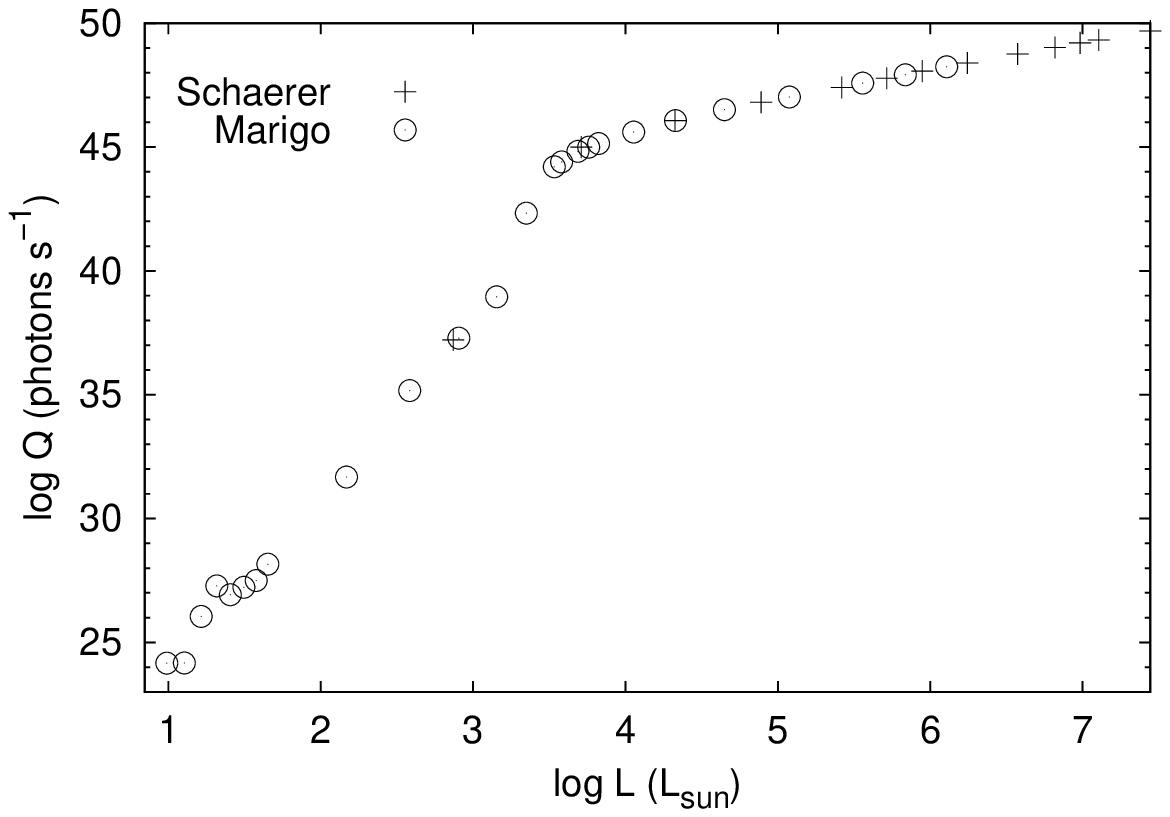}}
\caption{Number of ionizing photons $Q$ as a function of the stellar
luminosity for \ion{H}{i} (upper panel), \ion{He}{i} (middle panel), and
\ion{He}{ii} (lower panel) continua.
Stellar parameters are after \citet[][$+$]{ds02} and
\citet[][$\odot$]{zermetevol}.
}
\label{ionflux}
\end{figure}

\begin{figure}
\resizebox{\hsize}{!}{\includegraphics{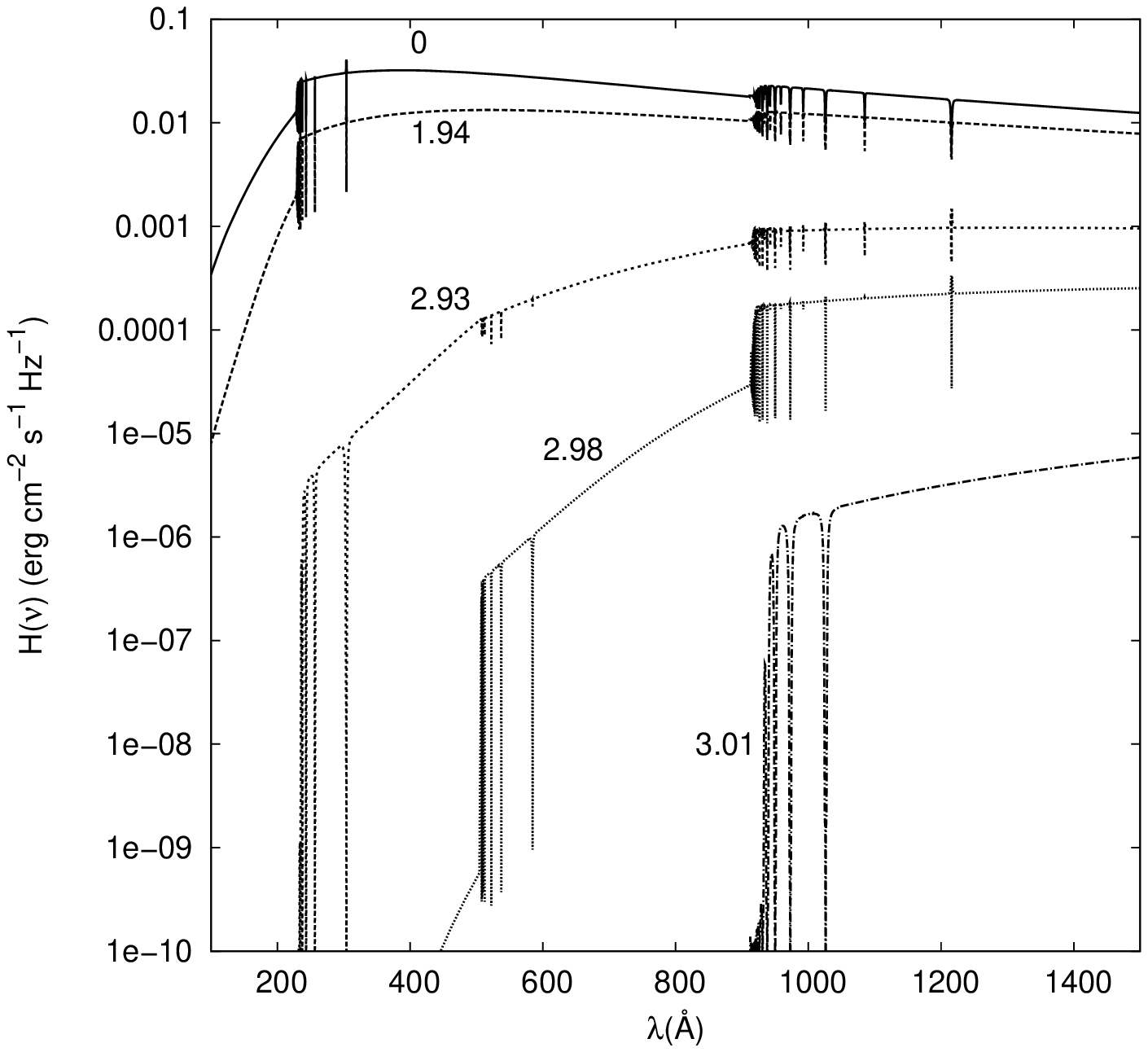}}
\resizebox{\hsize}{!}{\includegraphics{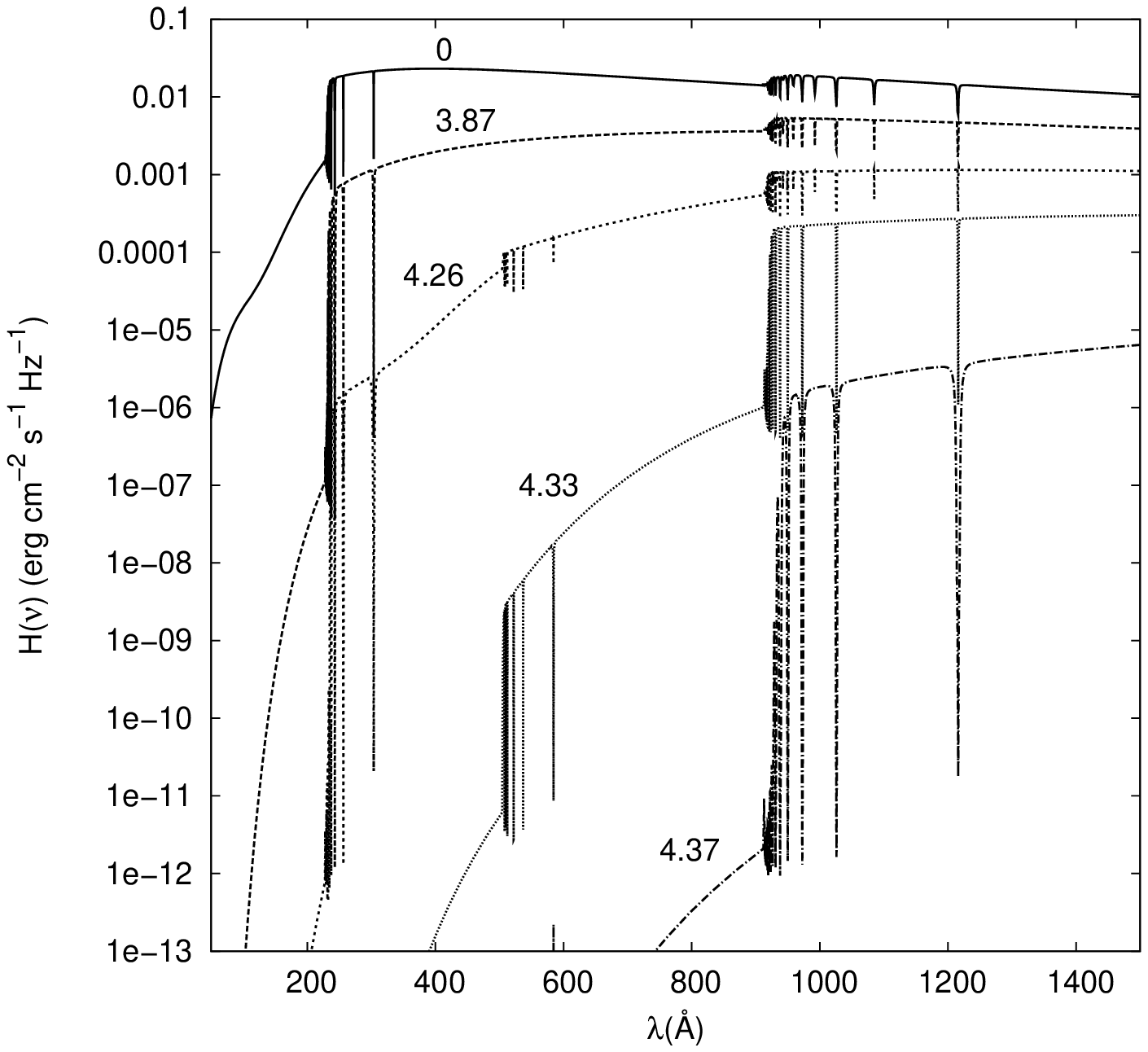}}
\caption{Plot of ionizing fluxes from selected model atmospheres after
\citet{zermetevol} for $100\Msun$ (upper panel) and $50\Msun$ (lower
panel) for different ages of first
stars with parameters given in Tables~\ref{qqevolv100} --
\ref{qqevolv10}..
The curves are labeled with the stellar age in megayears.}
\label{evolvflux}
\end{figure}

\begin{figure}
\resizebox{0.91\hsize}{!}{\includegraphics{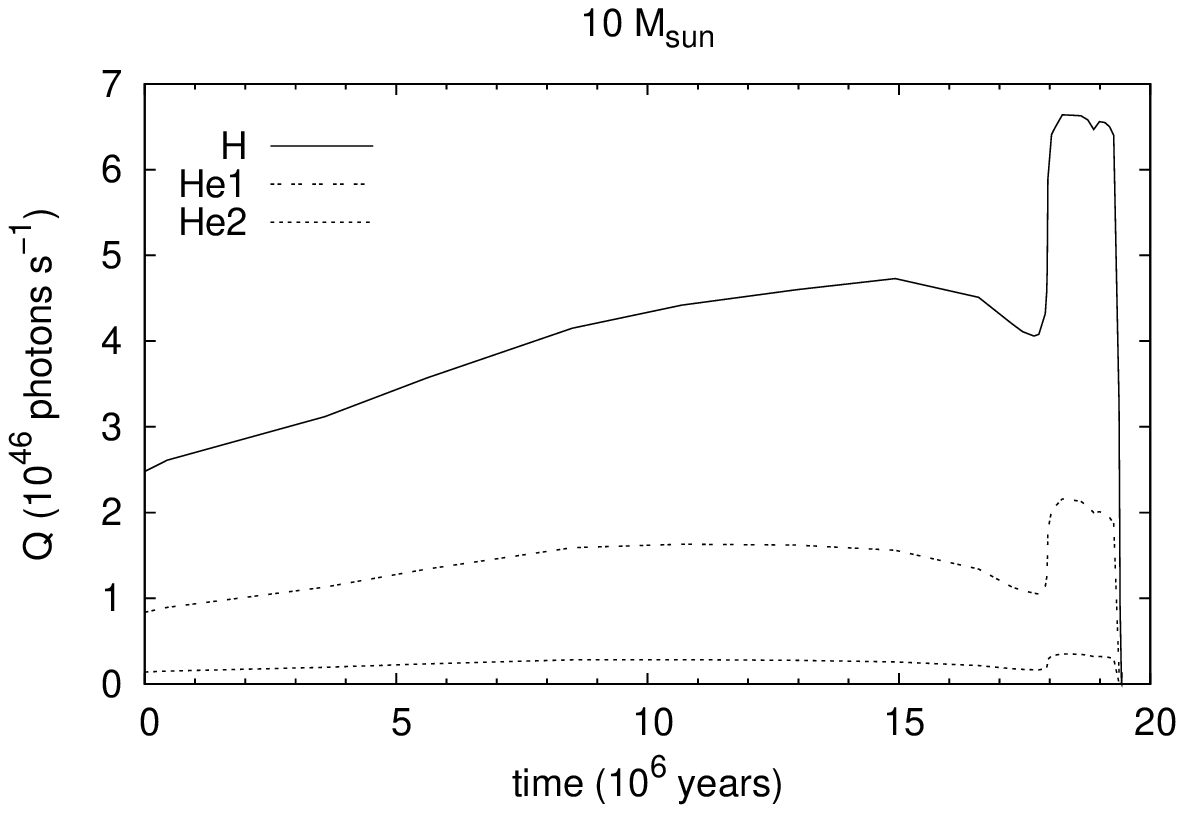}}
\resizebox{0.91\hsize}{!}{\includegraphics{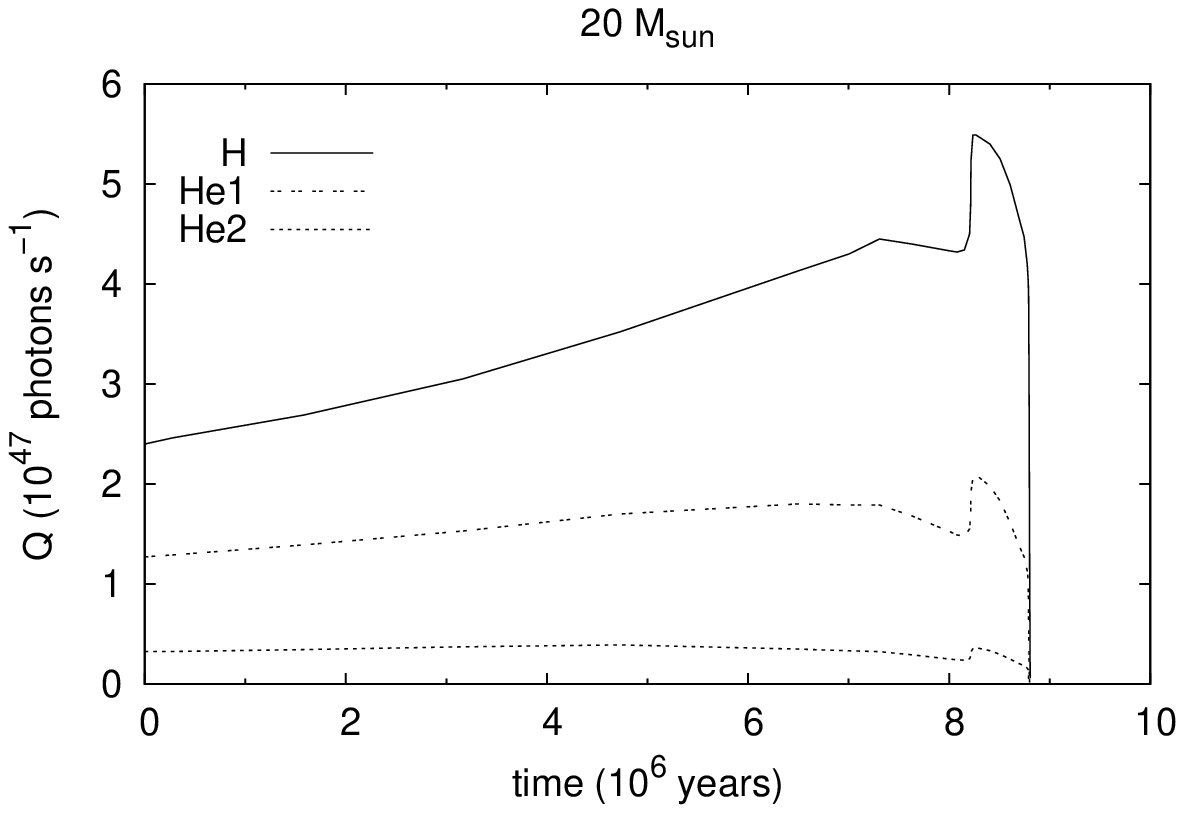}}
\resizebox{0.91\hsize}{!}{\includegraphics{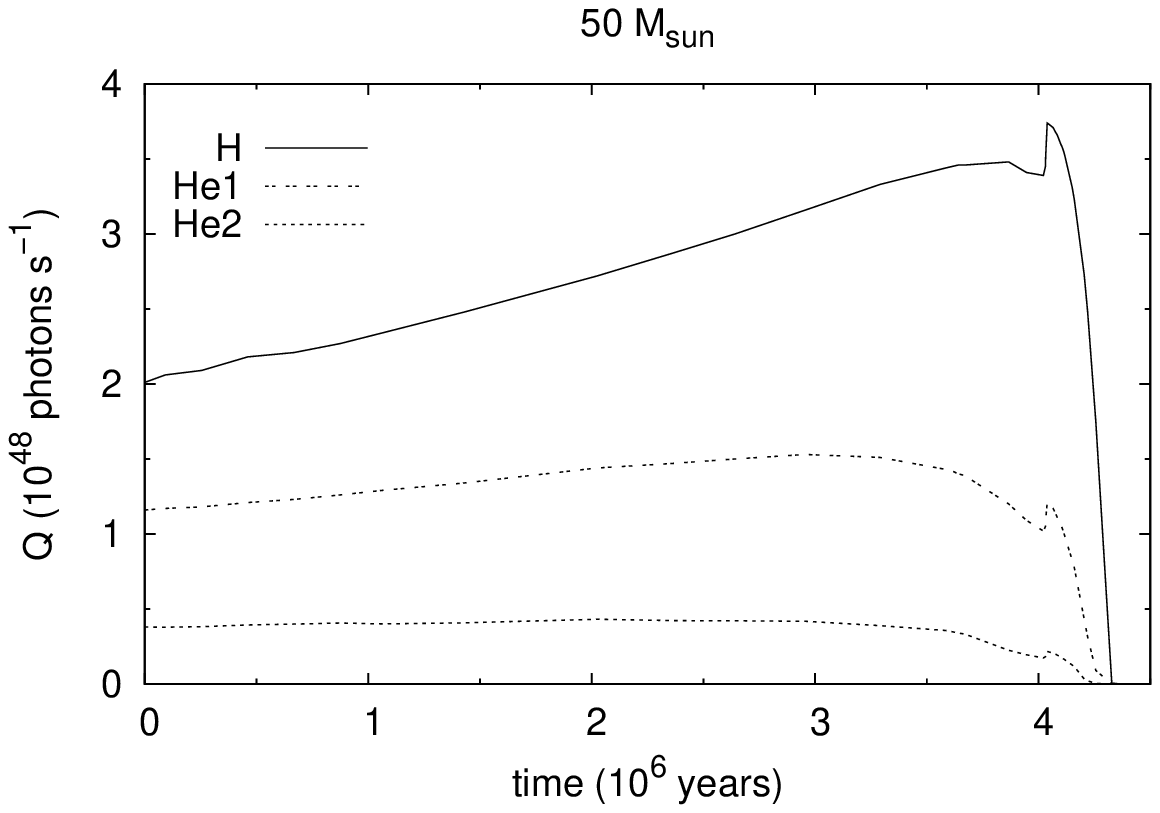}}
\resizebox{0.91\hsize}{!}{\includegraphics{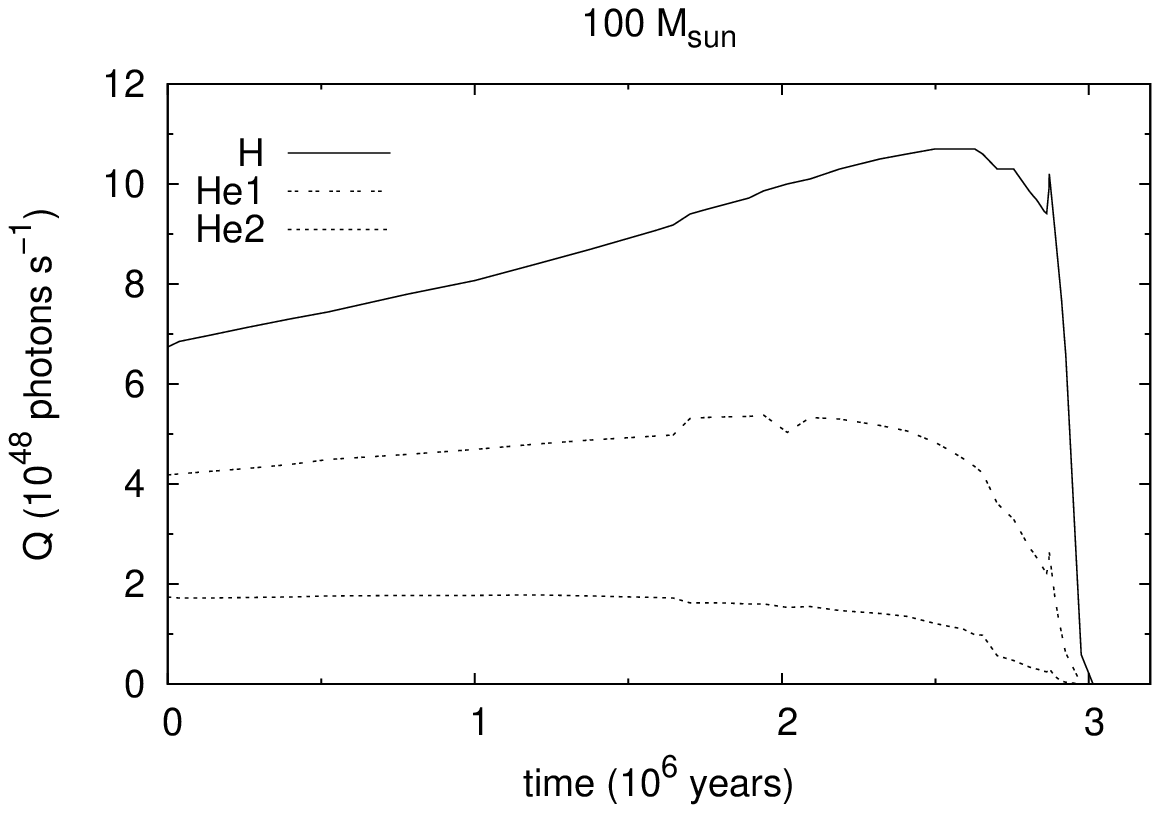}}
\caption{Number of ionizing photons $Q$ (Eq.~\ref{pocetfotonu}) in
\ion{H}{i}, \ion{He}{i}, and \ion{He}{ii} continua as a function of time
for different stellar masses ($10\Msun, 20\Msun, 50\Msun, 100\Msun$)
using evolutionary tracks from
\cite{zermetevol}.
The photon fluxes were calculated from model atmospheres calculated for
parameters given in the Tables~\ref{qqevolv100} -- \ref{qqevolv10}.}
\label{ioncas}
\end{figure}

\begin{figure}
\resizebox{\hsize}{!}{\includegraphics{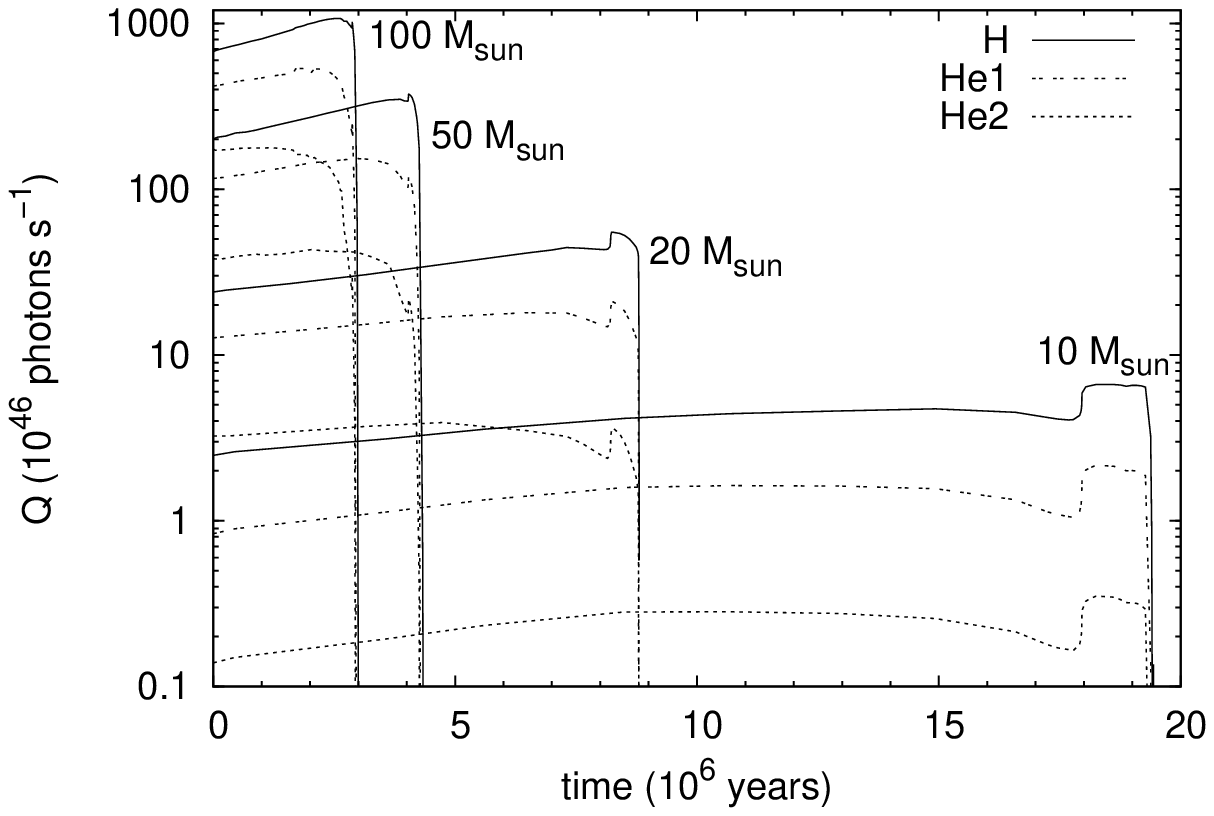}}
\caption{Comparison of the contribution to the ionizing photon flux from
Population III stars with different masses.
Model atmospheres are calculated for after \citet{zermetevol}.
The curves are labelled with the stellar mass.
The same data as for Fig.~\ref{ioncas} were used for this plot.}
\label{ionvse}
\end{figure}

As an immediate by-product of the model atmosphere calculation we can
obtain the amount of radiation escaping from the stellar photosphere.
This radiation significantly influences the circumstellar matter and it
may be the reason for stellar winds.
Fluxes from model atmospheres listed in the Table~\ref{qqzams} (see also
Fig.~\ref{zamstemp}) for selected parameters from the large
set of \citet{zermetevol} were used as a lower boundary condition in wind
calculations by \cite{cnovit} for analysis of the existence of stellar
winds.

For the case of Population III stars, the amount of highly energetic
emergent radiation is of extreme importance, because it influences the
ionization state of interstellar and intergalactic matter (therefore
this radiation is often referred to as the ionizing radiation) and,
consequently, it is important for the cosmic evolution
\citep{2003ApJ...594L...1V}.

Ultraviolet emergent fluxes from model atmospheres selected from the
lists in the Tables~\ref{qqds} and \ref{qqzams} are plotted in the
Figures~\ref{dsflux} and \ref{zamsflux}, respectively.
It is evident that the hottest and most massive (and hence most
luminous) stars provide maximum
flux in the far-ultraviolet region, which is the extremely important
spectral region for generation of the ``ionizing flux''.
Contribution to this flux from the low-massive zero-age main sequence
first stars is several orders of magnitude weaker.

For all models {\Iwe} calculate the stellar ionising photon flux $Q$ (in
photons s$^{-1}$) as
\begin{equation}\label{pocetfotonu}
Q = 4\pi R_\ast^2 \int_{\nu_\text{ion}}^\infty \frac{F_\nu}{h\nu} \dnu
\end{equation}
($\nu_\text{ion}$ is the ionization frequency of the particular
ionization edge) for \ion{H}{i},
\ion{He}{i}, and \ion{He}{ii} ionization edges.
The values of the stellar ionizing fluxes are listed in the
Tables~\ref{qqds} and \ref{qqzams}.
Comparing {\myour} results with those calculated by \cite{ds02} by a
totally independent code
(\TLUSTY)
we find a good agreement.
This supports the conclusion that both \citeauthor{ds02}'s and our
codes give correct results.

{\IWe} also plot the dependence of the number of ionizing photons for
\ion{H}{i}, \ion{He}{i}, and \ion{He}{ii} ionization edges on stellar
luminosity in the Figure~\ref{ionflux}.
As expected, the number of ionizing photons rises with the stellar
luminosity, the rise is significantly steeper for $L\lesssim3.5\Lsun$.

In addition, {\Iwe} studied the evolution of ionizing fluxes with the
stellar evolution.
For stellar masses 10\,{\Msun}, 20\,{\Msun}, 50\,{\Msun} and
1000\,{\Msun} {\Iwe} calculated model atmospheres for evolved Population
III stars.
Parameters of these models are listed in Tables \ref{qqevolv100} --
\ref{qqevolv10}.

The ultraviolet emergent fluxes of some of the calculated evolved model
atmospheres for masses 100\,{\Msun} and 50\,{\Msun} are shown in the
Figure~\ref{evolvflux}.
The latter figure shows that the high ionizing flux present at zero-age
main sequence first stars decreases with their evolution, as the stellar
effective temperature decreases.

In order to describe the changes of the ionizing flux with time
quantitatively, {\Iwe} calculated for each model listed in
Tables~\ref{qqevolv100} -- \ref{qqevolv10} the ionizing photon flux after
the Eq.~\eqref{pocetfotonu} for each ionization (\ion{H}{i},
\ion{He}{i}, \ion{He}{ii}), and {\Iwe} listed these numbers in
corresponding tables.
The changes of these fluxes with the stellar evolution are shown in
Figure \ref{ioncas}.
The fluxes first rise within the magnitude order of the initial value,
then after reaching the maximum value they quickly drop by several
orders of magnitude.
The maximum value flux depends on the stellar mass, it is reached sooner
for more massive stars.
The relative importance of first stars for the generation of ionizing
photons at different ages is shown in the Fig.~\ref{ionvse}.
The massive stars contibrute by several orders of magnitude more, but
since they evolve faster, their contribution lasts for much shorter time
period than for stars with lower mass.
For example, the $100\,\Msun$ star stops its contributing after
$3\cdot10^{6}$ years, while the $10\,\Msun$ star contributes for about
$19\cdot10^{6}$ years.
However, the total contribution of the  $10\,\Msun$ is smaller.

\section{Conclusions}

{\IWe} presented results of the NLTE model atmosphere calculations of
first stars (Population III stars) assuming hydrogen-helium composition.
These model atmospheres represent an extended set of model atmospheres
used as a lower boundary flux condition in wind analyses of first stars
by \citet{bezvi,cnovit}.
For the models {\Iwe} also calculated emergent fluxes and, in addition,
the ``ionizing photon fluxes'', which collect the contribution of the
radiation in the ultraviolet part of the spectrum.
Photon fluxes were calculated for radiation with shorter wavelength than
\ion{H}{i}, \ion{He}{i}, and \ion{He}{ii} ionization edges.
{\IWe} also studied the time evolution of all three photon fluxes and
{\Iwe} found strong dependence on the stellar mass, the more massive
stars are more important contributors to ionizing photon flux.
This underlines the importance of the question of the maximum stellar
mass for zero age first stars.

\acknowledgements
The author thanks Prof. Rolf-Peter Kudritzki for his comments to the
manuscript.
This research has made use of the NASA's Astrophysics Data System
Abstract Service.
This work was supported by a grant of the Grant Agency of the Czech
Republic
205/08/0003.
The Astronomical Institute Ond\v{r}ejov is supported by the project
RVO:67985815.

\nocite{*}
\bibliography{mp3,kubat,knihy}

\appendix

\section{Temperature structure of calculated model atmospheres}

\begin{figure}
\centering
\resizebox{!}{0.9\vsize}{\includegraphics{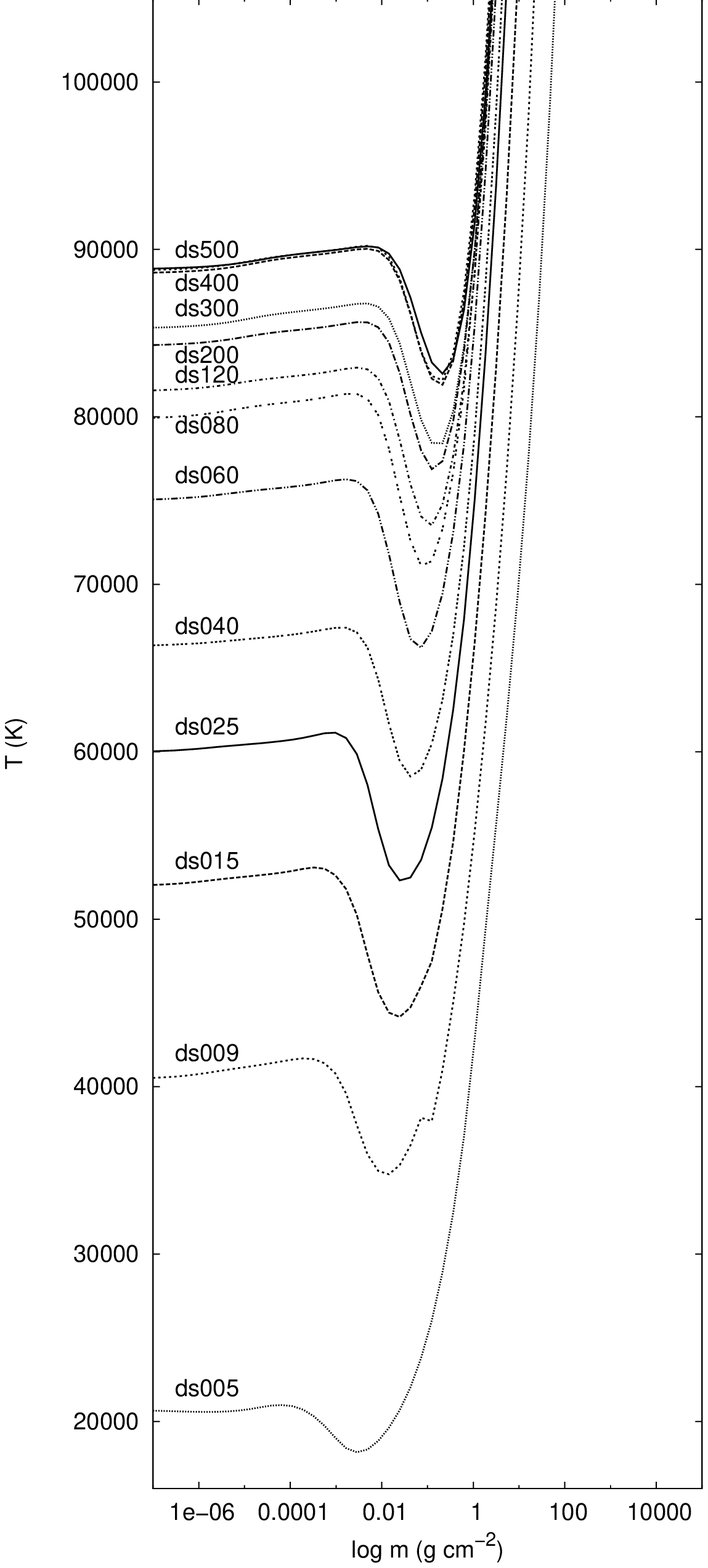}}
\caption{Run of temperature in NLTE model atmospheres of first stars
with parameters after \cite{ds02} listed in Table~\ref{qqds} (labels
correspond to the first column of the table).
The independent variable $m$ is the column mass-depth (see
Eq.~\ref{colmade}).
}
\label{dstemp}
\end{figure}

\begin{figure}
\centering
\resizebox{!}{0.96\vsize}{\includegraphics{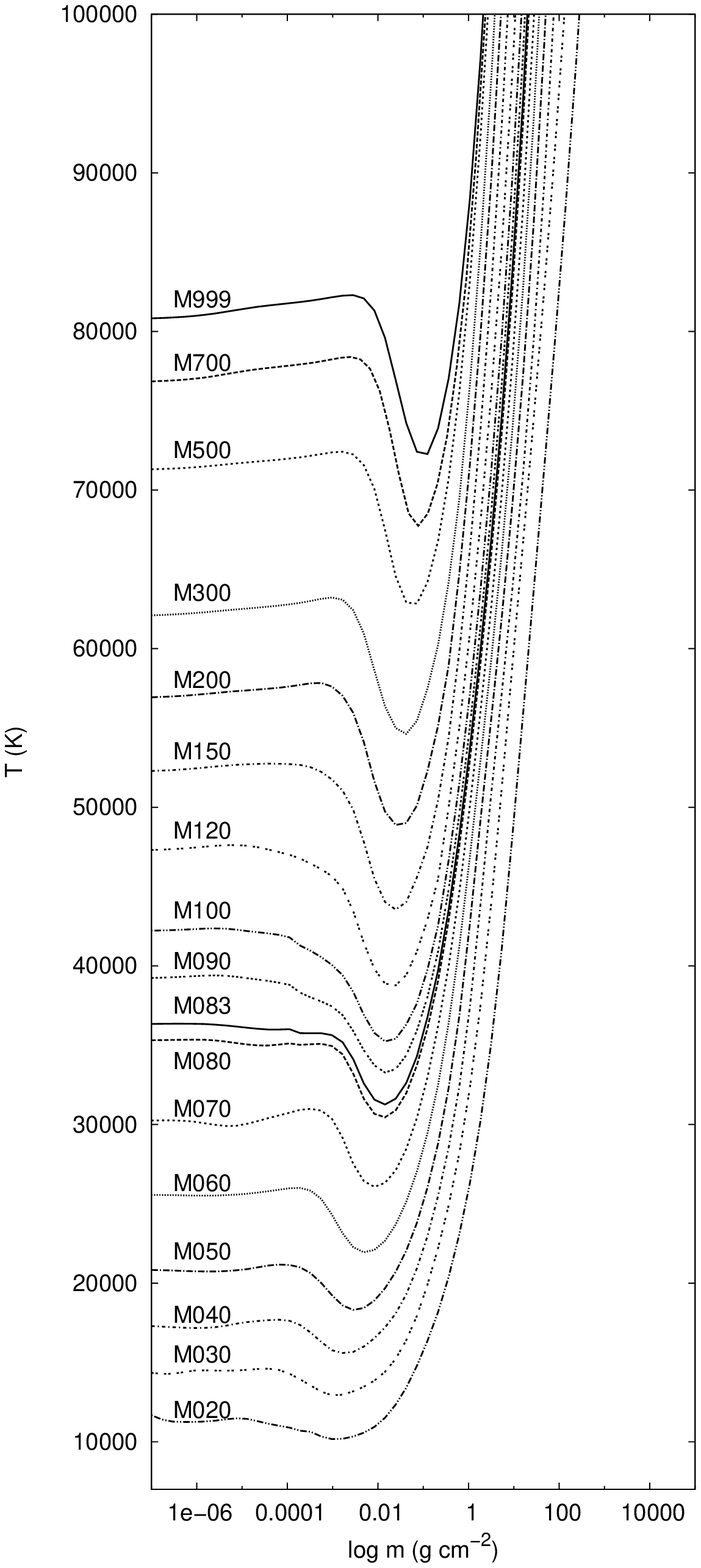}}
\caption{Run of temperature in NLTE model atmospheres of ZAMS first
stars with stellar parameters after
\cite{zermetevol}.
Individual labels Mxxx correspond to to the first column of the
Table~\ref{qqzams}.
The independent variable $m$ is the column mass-depth (see
Eq.~\ref{colmade}).
}
\label{zamstemp}
\end{figure}

\clearpage

\section{Ionizing fluxes during first star evolution}

\begin{table*}
\centering
\caption{Model atmosphere parameters and and stellar ionizing photon
fluxes $Q_i$ for selected model atmospheres from the evolutionary
sequence for $100\Msun$ for parameters from \cite{zermetevol}.}
\label{qqevolv100}
\begin{tabular}{lrlrrl|lll}
\hline
age & $R_*$ & $M_*$ & $L$ &
\Teff & $\log g$  
& $Q(\ion{H}{i})$
& $Q(\ion{He}{i})$
& $Q(\ion{He}{ii})$
\\
$[$years$]$ & [\Rsun] & [\Msun] & [\Lsun] & [\Kelvin] & &
\multicolumn{3}{c}{[\fotsek]} \\
\hline
0. & 4.23 & 100 & \dex{1.282}{6} & 94400 & 5.185 &
\dex{6.74}{48} & \dex{4.18}{48} & \dex{1.74}{48} \\
\dex{3.781}{4} & 4.43 & 100 & \dex{1.294}{6} & 92470 & 5.145 &
\dex{6.85}{48} & \dex{4.20}{48} & \dex{1.72}{48} \\
\dex{1.330}{5} & 4.54 & 100 & \dex{1.313}{6} & 91620 & 5.123 &
\dex{6.97}{48} & \dex{4.25}{48} & \dex{1.72}{48} \\
\dex{2.571}{5} & 4.69 & 100 & \dex{1.337}{6} & 90570 & 5.095 &
\dex{7.13}{48} & \dex{4.31}{48} & \dex{1.73}{48} \\
\dex{3.966}{5} & 4.83 & 100 & \dex{1.365}{6} & 89740 & 5.070 &
\dex{7.30}{48} & \dex{4.39}{48} & \dex{1.74}{48} \\
\dex{5.232}{5} & 4.96 & 100 & \dex{1.390}{6} & 88920 & 5.046 &
\dex{7.44}{48} & \dex{4.49}{48} & \dex{1.76}{48} \\
\dex{7.771}{5} & 5.28 & 100 & \dex{1.445}{6} & 87100 & 4.993 &
\dex{7.79}{48} & \dex{4.59}{48} & \dex{1.77}{48} \\
\dex{1.001}{6} & 5.57 & 100 & \dex{1.496}{6} & 85500 & 4.946 &
\dex{8.07}{48} & \dex{4.69}{48} & \dex{1.77}{48} \\
\dex{1.201 }{6} & 5.91 & 100 & \dex{1.549}{6} & 83750 & 4.895
& \dex{8.40}{48} & \dex{4.80}{48} & \dex{1.78}{48} \\
\dex{1.375}{6} & 6.27 & 100 & \dex{1.592}{6} & 81840 & 4.843
& \dex{8.69}{48} & \dex{4.88}{48} & \dex{1.76}{48} \\
\dex{1.589}{6} & 6.81 & 100 & \dex{1.652}{6} & 79250 & 4.771
& \dex{9.07}{48} & \dex{4.96}{48} & \dex{1.73}{48} \\
\dex{1.647}{6} & 7.01 & 100 & \dex{1.671}{6} & 78340 & 4.746
& \dex{9.18}{48} & \dex{4.98}{48} & \dex{1.72}{48} \\
\dex{1.701}{6} & 7.18 & 100 & \dex{1.687}{6} & 77620 & 4.726
& \dex{9.40}{48} & \dex{5.31}{48} & \dex{1.62}{48} \\
\dex{1.753}{6} & 7.34 & 100 & \dex{1.702}{6} & 76910 & 4.706
& \dex{9.49}{48} & \dex{5.33}{48} & \dex{1.62}{48} \\
\dex{1.802}{6} & 7.55 & 100 & \dex{1.718}{6} & 76030 & 4.682
& \dex{9.57}{48} & \dex{5.34}{48} & \dex{1.62}{48} \\
\dex{1.893}{6} & 7.94 & 100 & \dex{1.745}{6} & 74470 & 4.639
& \dex{9.72}{48} & \dex{5.35}{48} & \dex{1.60}{48} \\
\dex{1.941}{6} & 8.18 & 100 & \dex{1.773}{6} & 73620 & 4.612
& \dex{9.86}{48} & \dex{5.38}{48} & \dex{1.60}{48} \\
\dex{2.018}{6} & 8.57 & 100 & \dex{1.791}{6} & 72110 & 4.572
& \dex{1.00}{49} & \dex{5.03}{48} & \dex{1.53}{48} \\
\dex{2.093}{6} & 9.03 & 100 & \dex{1.816}{6} & 70470 & 4.526
& \dex{1.01}{49} & \dex{5.33}{48} & \dex{1.55}{48} \\
\dex{2.188}{6} & 9.72 & 100 & \dex{1.849}{6} & 68230 & 4.462
& \dex{1.03}{49} & \dex{5.30}{48} & \dex{1.47}{48} \\
\dex{2.319}{6} & 11.0 & 100 & \dex{1.897}{6} & 64560 & 4.355
& \dex{1.05}{49} & \dex{5.17}{48} & \dex{1.41}{48} \\
\dex{2.408}{6} & 12.1 & 100 & \dex{1.932}{6} & 61940 & 4.275
& \dex{1.06}{49} & \dex{5.06}{48} & \dex{1.35}{48} \\
\dex{2.500}{6} & 13.7 & 100 & \dex{1.968}{6} & 58480 & 4.167
& \dex{1.07}{49} & \dex{4.83}{48} & \dex{1.21}{48} \\
\dex{2.547}{6} & 14.6 & 100 & \dex{1.986}{6} & 56620 & 4.107
& \dex{1.07}{49} & \dex{4.67}{48} & \dex{1.15}{48} \\
\dex{2.590}{6} & 15.7 & 100 & \dex{2.004}{6} & 54830 & 4.047
& \dex{1.07}{49} & \dex{4.51}{48} & \dex{1.10}{48} \\
\dex{2.603}{6} & 16.1 & 100 & \dex{2.009}{6} & 54200 & 4.026
& \dex{1.07}{49} & \dex{4.45}{48} & \dex{1.06}{48} \\
\dex{2.628}{6} & 16.8 & 100 & \dex{2.023}{6} & 53090 & 3.987
& \dex{1.07}{49} & \dex{4.35}{48} & \dex{9.85}{47} \\
\dex{2.654}{6} & 17.6 & 100 & \dex{2.032}{6} & 51880 & 3.945
& \dex{1.06}{49} & \dex{4.22}{48} & \dex{9.74}{47} \\
\dex{2.701}{6} & 19.6 & 100 & \dex{2.056}{6} & 49320 & 3.852
& \dex{1.03}{49} & \dex{3.61}{48} & \dex{5.67}{47} \\
\dex{2.755}{6} & 22.5 & 100 & \dex{2.080}{6} & 46240 & 3.735
& \dex{1.03}{49} & \dex{3.29}{48} & \dex{4.70}{47} \\
\dex{2.805}{6} & 26.0 & 100 & \dex{2.109}{6} & 43150 & 3.609
& \dex{9.86}{48} & \dex{2.74}{48} & \dex{3.41}{47} \\
\dex{2.812}{6} & 26.6 & 100 & \dex{2.109}{6} & 42660 & 3.589
& \dex{9.81}{48} & \dex{2.68}{48} & \dex{3.28}{47} \\
\dex{2.820}{6} & 27.2 & 100 & \dex{2.113}{6} & 42170 & 3.568
& \dex{9.75}{48} & \dex{2.62}{48} & \dex{3.17}{47} \\
\dex{2.827}{6} & 27.9 & 100 & \dex{2.118}{6} & 41690 & 3.547
& \dex{9.70}{48} & \dex{2.55}{48} & \dex{3.03}{47} \\
\dex{2.834}{6} & 28.6 & 100 & \dex{2.123}{6} & 41210 & 3.526
& \dex{9.64}{48} & \dex{2.49}{48} & \dex{2.91}{47} \\
\dex{2.855}{6} & 30.7 & 100 & \dex{2.138}{6} & 39810 & 3.463
& \dex{9.45}{48} & \dex{2.28}{48} & \dex{2.54}{47} \\
\dex{2.863}{6} & 31.5 & 100 & \dex{2.148}{6} & 39350 & 3.441
& \dex{9.41}{48} & \dex{2.22}{48} & \dex{2.44}{47} \\
\dex{2.870}{6} & 30.0 & 100 & \dex{2.193}{6} & 40550 & 3.484
& \dex{9.90}{48} & \dex{2.46}{48} & \dex{2.75}{47} \\
\dex{2.870}{6} & 29.4 & 100 & \dex{2.203}{6} & 41020 & 3.502
& \dex{1.01}{49} & \dex{2.56}{48} & \dex{2.90}{47} \\
\dex{2.871}{6} & 28.9 & 100 & \dex{2.213}{6} & 41400 & 3.516
& \dex{1.02}{49} & \dex{2.63}{48} & \dex{3.02}{47} \\
\dex{2.873}{6} & 29.6 & 100 & \dex{2.218}{6} & 40930 & 3.495
& \dex{1.01}{49} & \dex{2.56}{48} & \dex{2.91}{47} \\
\dex{2.874}{6} & 30.3 & 100 & \dex{2.218}{6} & 40460 & 3.475
& \dex{1.00}{49} & \dex{2.47}{48} & \dex{2.77}{47} \\
\dex{2.889}{6} & 37.4 & 100 & \dex{2.239}{6} & 36470 & 3.291
& \dex{9.09}{48} & \dex{1.74}{48} & \dex{1.55}{47} \\
\dex{2.900}{6} & 42.5 & 100 & \dex{2.249}{6} & 34280 & 3.181
& \dex{8.40}{48} & \dex{1.35}{48} & \dex{1.05}{47} \\
\dex{2.911}{6} & 48.0 & 100 & \dex{2.259}{6} & 32280 & 3.075
& \dex{7.67}{48} & \dex{1.01}{48} & \dex{6.74}{46} \\
\dex{2.920}{6} & 53.4 & 100 & \dex{2.265}{6} & 30620 & 2.982
& \dex{6.93}{48} & \dex{7.31}{47} & \dex{4.25}{46} \\
\dex{2.925}{6} & 56.4 & 100 & \dex{2.280}{6} & 29850 & 2.935
& \dex{6.58}{48} & \dex{6.18}{47} & \dex{3.35}{46 } \\
\dex{2.975}{6} & 125. & 100 & \dex{2.311}{6} & 20090 & 2.241
& \dex{5.86}{47} & \dex{6.01}{42} & \dex{3.15}{39} \\
\dex{3.014}{6} & 510. & 100 & \dex{2.327}{6} &  9977 & 1.022
& \dex{3.86}{42} & \dex{1.05}{36} & \dex{1.32}{27} \\
\hline
\end{tabular}
\end{table*}

\begin{table*}
\centering
\caption{The same as Table~\ref{qqevolv100} but for $50\Msun$.}
\label{qqevolv50}
\begin{tabular}{lrlrll|lll}
\hline
age & $R_*$ & $M_*$ & $L$ &
\Teff & $\log g$  
& $Q(\ion{H}{i})$
& $Q(\ion{He}{i})$
& $Q(\ion{He}{ii})$
\\
$[$years$]$ & [\Rsun] & [\Msun] & [\Lsun] & [\Kelvin] & &
\multicolumn{3}{c}{[\fotsek]} \\
\hline
0             & 2.82 & 50 & \dex{3.597}{5} & 84140 & 5.236 &
\dex{2.01}{48} & \dex{1.16}{48} & \dex{3.80}{47} \\
\dex{9.203}{4} & 2.99 & 50 & \dex{3.664}{5} & 82035 & 5.183 &
\dex{2.06}{48} & \dex{1.17}{48} & \dex{3.79}{47} \\
\dex{2.545}{5} & 3.08 & 50 & \dex{3.741}{5} & 81280 & 5.159 &
\dex{2.09}{48} & \dex{1.18}{48} & \dex{3.82}{47} \\
\dex{4.606}{5} & 3.17 & 50 & \dex{3.846}{5} & 80720 & 5.135 &
\dex{2.18}{48} & \dex{1.21}{48} & \dex{3.93}{47} \\
\dex{6.683}{5} & 3.27 & 50 & \dex{3.954}{5} & 79980 & 5.107 &
\dex{2.21}{48} & \dex{1.23}{48} & \dex{3.99}{47} \\
\dex{8.760}{5} & 3.38 & 50 & \dex{4.064}{5} & 79250 & 5.079 &
\dex{2.27}{48} & \dex{1.26}{48} & \dex{4.06}{47} \\
\dex{1.063}{6} & 3.47 & 50 & \dex{4.178}{5} & 78700 & 5.055 &
\dex{2.34}{48} & \dex{1.29}{48} & \dex{4.00}{47} \\
\dex{1.432}{6} & 3.71 & 50 & \dex{4.416}{5} & 77270 & 4.999 &
\dex{2.48}{48} & \dex{1.34}{48} & \dex{4.07}{47} \\
\dex{2.024}{6} & 4.18 & 50 & \dex{4.842}{5} & 74470 & 4.895 &
\dex{2.72}{48} & \dex{1.44}{48} & \dex{4.32}{47} \\
\dex{2.359}{6} & 4.56 & 50 & \dex{5.129}{5} & 72280 & 4.818 &
\dex{2.87}{48} & \dex{1.47}{48} & \dex{4.23}{47} \\
\dex{2.642}{6} & 4.99 & 50 & \dex{5.383}{5} & 69980 & 4.741 &
\dex{3.00}{48} & \dex{1.50}{48} & \dex{4.21}{47} \\
\dex{2.957}{6} & 5.63 & 50 & \dex{5.702}{5} & 66830 & 4.636 &
\dex{3.16}{48} & \dex{1.53}{48} & \dex{4.18}{47} \\
\dex{3.289}{6} & 6.68 & 50 & \dex{6.095}{5} & 62370 & 4.487 &
\dex{3.33}{48} & \dex{1.51}{48} & \dex{3.89}{47} \\
\dex{3.584}{6} & 8.20 & 50 & \dex{6.471}{5} & 57150 & 4.309 &
\dex{3.44}{48} & \dex{1.43}{48} & \dex{3.56}{47} \\
\dex{3.613}{6} & 8.42 & 50 & \dex{6.516}{5} & 56490 & 4.286 &
\dex{3.45}{48} & \dex{1.42}{48} & \dex{3.48}{47} \\
\dex{3.640}{6} & 8.65 & 50 & \dex{6.561}{5} & 55850 & 4.263 &
\dex{3.46}{48} & \dex{1.40}{48} & \dex{3.39}{47} \\
\dex{3.669}{6} & 8.87 & 50 & \dex{6.592}{5} & 55210 & 4.241 &
\dex{3.46}{48} & \dex{1.39}{48} & \dex{3.33}{47} \\
\dex{3.866}{6} & 11.1 & 50 & \dex{6.949}{5} & 50000 & 4.046 &
\dex{3.48}{48} & \dex{1.20}{48} & \dex{2.24}{47} \\
\dex{3.946}{6} & 12.4 & 50 & \dex{7.046}{5} & 47530 & 3.952 &
\dex{3.41}{48} & \dex{1.09}{48} & \dex{1.95}{47} \\
\dex{4.021}{6} & 13.6 & 50 & \dex{7.228}{5} & 45600 & 3.869 &
\dex{3.39}{48} & \dex{1.02}{48} & \dex{1.74}{47} \\
\dex{4.031}{6} & 13.4 & 50 & \dex{7.311}{5} & 46020 & 3.880 &
\dex{3.45}{48} & \dex{1.05}{48} & \dex{1.82}{47} \\
\dex{4.032}{6} & 13.2 & 50 & \dex{7.362}{5} & 46560 & 3.897 &
\dex{3.51}{48} & \dex{1.09}{48} & \dex{1.90}{47} \\
\dex{4.039}{6} & 12.6 & 50 & \dex{7.621}{5} & 48080 & 3.938 &
\dex{3.74}{48} & \dex{1.21}{48} & \dex{2.18}{47} \\
\dex{4.047}{6} & 12.8 & 50 & \dex{7.656}{5} & 47640 & 3.920 &
\dex{3.73}{48} & \dex{1.19}{48} & \dex{2.13}{47} \\
\dex{4.064}{6} & 13.1 & 50 & \dex{7.673}{5} & 47100 & 3.899 &
\dex{3.71}{48} & \dex{1.17}{48} & \dex{2.06}{47} \\
\dex{4.075}{6} & 13.5 & 50 & \dex{7.691}{5} & 46560 & 3.878 &
\dex{3.68}{48} & \dex{1.14}{48} & \dex{1.99}{47} \\
\dex{4.084}{6} & 13.8 & 50 & \dex{7.709}{5} & 46020 & 3.857 &
\dex{3.66}{48} & \dex{1.11}{48} & \dex{1.92}{47} \\
\dex{4.092}{6} & 14.1 & 50 & \dex{7.727}{5} & 45500 & 3.836 &
\dex{3.63}{48} & \dex{1.09}{48} & \dex{1.86}{47} \\
\dex{4.099}{6} & 14.5 & 50 & \dex{7.727}{5} & 44980 & 3.816 &
\dex{3.60}{48} & \dex{1.06}{48} & \dex{1.78}{47} \\
\dex{4.106}{6} & 14.8 & 50 & \dex{7.745}{5} & 44560 & 3.799 &
\dex{3.58}{48} & \dex{1.04}{48} & \dex{1.73}{47} \\
\dex{4.112}{6} & 15.1 & 50 & \dex{7.762}{5} & 44050 & 3.778 &
\dex{3.55}{48} & \dex{1.01}{48} & \dex{1.67}{47} \\
\dex{4.118}{6} & 15.5 & 50 & \dex{7.780}{5} & 43550 & 3.757 &
\dex{3.52}{48} & \dex{9.83}{47} & \dex{1.61}{47} \\
\dex{4.150}{6} & 17.8 & 50 & \dex{7.852}{5} & 40740 & 3.637 &
\dex{3.31}{48} & \dex{8.25}{47} & \dex{1.26}{47} \\
\dex{4.159}{6} & 18.6 & 50 & \dex{7.870}{5} & 39900 & 3.600 &
\dex{3.23}{48} & \dex{7.77}{47} & \dex{1.16}{47} \\
\dex{4.204}{6} & 23.6 & 50 & \dex{7.980}{5} & 35480 & 3.390 &
\dex{2.73}{48} & \dex{4.33}{47} & \dex{3.37}{46} \\
\dex{4.220}{6} & 26.1 & 50 & \dex{8.017}{5} & 33810 & 3.304 &
\dex{2.48}{48} & \dex{3.14}{47} & \dex{2.11}{46} \\
\dex{4.256}{6} & 33.7 & 50 & \dex{8.139}{5} & 29850 & 3.081 &
\dex{1.75}{48} & \dex{9.08}{46} & \dex{3.89}{45} \\
\dex{4.328}{6} & 72.0 & 50 & \dex{8.355}{5} & 20560 & 2.422 &
\dex{6.34}{45} & \dex{2.04}{40} & \dex{3.49}{35} \\
\dex{4.366}{6} & 303. & 50 & \dex{8.502}{5} & 10070 & 1.174 &
\dex{1.29}{41} & \dex{5.33}{35} & \dex{2.07}{26} \\
\hline
\end{tabular}
\end{table*}

\begin{table*}
\centering
\caption{The same as Table~\ref{qqevolv100} but for $20\Msun$.}
\label{qqevolv20}
\begin{tabular}{lrlrll|lll}
\hline
age & $R_*$ & $M_*$ & $L$ &
\Teff & $\log g$  
& $Q(\ion{H}{i})$
& $Q(\ion{He}{i})$
& $Q(\ion{He}{ii})$
\\
$[$years$]$ & [\Rsun] & [\Msun] & [\Lsun] & [\Kelvin] & &
\multicolumn{3}{c}{[\fotsek]} \\
\hline

0              & 1.65 & 20 & \dex{4.456}{4} & 65310 & 5.305 &
\dex{2.40}{47} & \dex{1.27}{47} & \dex{3.24}{46} \\
\dex{2.720}{5} & 1.73 & 20 & \dex{4.613}{4} & 64120 & 5.258 &
\dex{2.46}{47} & \dex{1.29}{47} & \dex{3.24}{46 } \\
\dex{1.581}{6} & 1.90 & 20 & \dex{5.093}{4} & 62800 & 5.179 &
\dex{2.69}{47} & \dex{1.39}{47} & \dex{3.44}{46} \\
\dex{3.164}{6} & 2.15 & 20 & \dex{5.834}{4} & 61090 & 5.072 &
\dex{3.05}{47} & \dex{1.53}{47} & \dex{3.71}{46} \\
\dex{4.726}{6} & 2.52 & 20 & \dex{6.776}{4} & 58610 & 4.935 &
\dex{3.52}{47} & \dex{1.70}{47} & \dex{3.91}{46} \\
\dex{6.497}{6} & 3.31 & 20 & \dex{8.317}{4} & 53830 & 4.698 &
\dex{4.13}{47} & \dex{1.80}{47} & \dex{3.49}{46} \\
\dex{7.004}{6} & 3.73 & 20 & \dex{8.913}{4} & 51640 & 4.596 &
\dex{4.30}{47} & \dex{1.79}{47} & \dex{3.32}{46} \\
\dex{7.306}{6} & 4.10 & 20 & \dex{9.480}{4} & 50000 & 4.513 &
\dex{4.45}{47} & \dex{1.79}{47} & \dex{3.23}{46} \\
\dex{7.630}{6} & 4.54 & 20 & \dex{9.772}{4} & 47860 & 4.424 &
\dex{4.40}{47} & \dex{1.68}{47} & \dex{2.92}{46} \\
\dex{8.012}{6} & 5.34 & 20 & \dex{1.042}{5} & 44870 & 4.284 &
\dex{4.33}{47} & \dex{1.52}{47} & \dex{2.49}{46} \\
\dex{8.078}{6} & 5.50 & 20 & \dex{1.057}{5} & 44360 & 4.258 &
\dex{4.32}{47} & \dex{1.49}{47} & \dex{2.42}{46} \\
\dex{8.151}{6} & 5.65 & 20 & \dex{1.074}{5} & 43950 & 4.235 &
\dex{4.34}{47} & \dex{1.48}{47} & \dex{2.38}{46} \\
\dex{8.202}{6} & 5.60 & 20 & \dex{1.096}{5} & 44360 & 4.242 &
\dex{4.50}{47} & \dex{1.55}{47} & \dex{2.51}{46} \\
\dex{8.212}{6} & 5.46 & 20 & \dex{1.132}{5} & 45290 & 4.264 &
\dex{4.78}{47} & \dex{1.69}{47} & \dex{2.78}{46} \\
\dex{8.216}{6} & 5.28 & 20 & \dex{1.191}{5} & 46660 & 4.294 &
\dex{5.23}{47} & \dex{1.92}{47} & \dex{3.22}{46} \\
\dex{8.235}{6} & 5.09 & 20 & \dex{1.216}{5} & 47750 & 4.325 &
\dex{5.49}{47} & \dex{2.06}{47} & \dex{3.53}{46} \\
\dex{8.267}{6} & 4.94 & 20 & \dex{1.202}{5} & 48300 & 4.350 &
\dex{5.49}{47} & \dex{2.09}{47} & \dex{3.62}{46} \\
\dex{8.405}{6} & 5.36 & 20 & \dex{1.225}{5} & 46770 & 4.286 &
\dex{5.40}{47} & \dex{1.98}{47} & \dex{3.33}{46} \\
\dex{8.506}{6} & 5.85 & 20 & \dex{1.250}{5} & 44870 & 4.205 &
\dex{5.25}{47} & \dex{1.83}{47} & \dex{2.95}{46} \\
\dex{8.607}{6} & 6.57 & 20 & \dex{1.279}{5} & 42560 & 4.103 &
\dex{4.99}{47} & \dex{1.61}{47} & \dex{2.49}{46} \\
\dex{8.697}{6} & 7.44 & 20 & \dex{1.312}{5} & 40270 & 3.996 &
\dex{4.65}{47} & \dex{1.38}{47} & \dex{2.03}{46} \\
\dex{8.744}{6} & 8.00 & 20 & \dex{1.334}{5} & 38990 & 3.933 &
\dex{4.47}{47} & \dex{1.27}{47} & \dex{1.82}{46} \\
\dex{8.774}{6} & 8.58 & 20 & \dex{1.349}{5} & 37760 & 3.872 &
\dex{4.20}{47} & \dex{1.13}{47} & \dex{1.58}{46} \\
\dex{8.786}{6} & 8.97 & 20 & \dex{1.358}{5} & 36980 & 3.833 &
\dex{4.02}{47} & \dex{1.04}{47} & \dex{1.43}{46} \\
\dex{8.790}{6} & 9.18 & 20 & \dex{1.358}{5} & 36560 & 3.813 &
\dex{3.87}{47} & \dex{5.84}{46} & \dex{5.18}{45} \\
\dex{8.795}{6} & 10.5 & 20 & \dex{1.312}{5} & 33960 & 3.700 &
\dex{2.81}{47} & \dex{1.97}{46} & \dex{1.25}{45} \\
\dex{8.800}{6} & 13.8 & 20 & \dex{1.318}{5} & 29580 & 3.458 &
\dex{1.11}{47} & \dex{1.68}{44} & \dex{1.58}{41} \\
\dex{8.805}{6} & 19.9 & 20 & \dex{1.297}{5} & 24550 & 3.141 &
\dex{5.78}{45} & \dex{7.50}{41} & \dex{1.91}{38} \\

\hline
\end{tabular}
\end{table*}

\begin{table*}
\centering
\caption{The same as Table~\ref{qqevolv100} but for $10\Msun$.}
\label{qqevolv10}
\begin{tabular}{lrlrrl|lll}
\hline
age & $R_*$ & $M_*$ & $L$ &
\Teff & $\log g$  
& $Q(\ion{H}{i})$
& $Q(\ion{He}{i})$
& $Q(\ion{He}{ii})$
\\
$[$years$]$ & [\Rsun] & [\Msun] & [\Lsun] & [\Kelvin] & &
\multicolumn{3}{c}{[\fotsek]} \\
\hline
 0             & 1.37 & 10 & \dex{6.653}{3} & 44460 & 5.162 &
\dex{2.48}{46} & \dex{8.36}{45} & \dex{1.39}{45} \\
\dex{4.445}{5} & 1.38 & 10 & \dex{6.887}{3} & 44770 & 5.159 &
\dex{2.61}{46} & \dex{8.94}{45} & \dex{1.50}{45} \\
\dex{3.584}{6} & 1.39 & 10 & \dex{7.852}{3} & 46020 & 5.150 &
\dex{3.12}{46} & \dex{1.13}{46} & \dex{1.94}{45} \\
\dex{5.616}{6} & 1.41 & 10 & \dex{8.690}{3} & 46990 & 5.142 &
\dex{3.57}{46} & \dex{1.34}{46} & \dex{2.34}{45} \\
\dex{8.502}{6} & 1.47 & 10 & \dex{9.954}{3} & 47530 & 5.103 &
\dex{4.15}{46} & \dex{1.59}{46} & \dex{2.82}{45} \\
\dex{1.068}{7} & 1.62 & 10 & \dex{1.081}{4} & 46240 & 5.019 &
\dex{4.42}{46} & \dex{1.63}{46} & \dex{2.83}{45} \\
\dex{1.296}{7} & 1.83 & 10 & \dex{1.197}{4} & 44560 & 4.911 &
\dex{4.60}{46} & \dex{1.62}{46} & \dex{2.76}{45} \\
\dex{1.493}{7} & 2.13 & 10 & \dex{1.327}{4} & 42460 & 4.782 &
\dex{4.73}{46} & \dex{1.56}{46} & \dex{2.57}{45} \\
\dex{1.658}{7} & 2.52 & 10 & \dex{1.469}{4} & 39990 & 4.634 &
\dex{4.51}{46} & \dex{1.34}{46} & \dex{2.14}{45} \\
\dex{1.725}{7} & 2.76 & 10 & \dex{1.541}{4} & 38720 & 4.557 &
\dex{4.20}{46} & \dex{1.13}{46} & \dex{1.79}{45} \\
\dex{1.746}{7} & 2.84 & 10 & \dex{1.567}{4} & 38280 & 4.530 &
\dex{4.11}{46} & \dex{1.09}{46} & \dex{1.71}{45} \\
\dex{1.769}{7} & 2.92 & 10 & \dex{1.596}{4} & 37930 & 4.506 &
\dex{4.06}{46} & \dex{1.06}{46} & \dex{1.67}{45} \\
\dex{1.778}{7} & 2.95 & 10 & \dex{1.614}{4} & 37840 & 4.497 &
\dex{4.08}{46} & \dex{1.05}{46} & \dex{1.65}{45} \\
\dex{1.791}{7} & 2.93 & 10 & \dex{1.652}{4} & 38190 & 4.503 &
\dex{4.32}{46} & \dex{1.14}{46} & \dex{1.80}{45} \\
\dex{1.794}{7} & 2.89 & 10 & \dex{1.679}{4} & 38640 & 4.516 &
\dex{4.57}{46} & \dex{1.24}{46} & \dex{1.96}{45} \\
\dex{1.795}{7} & 2.87 & 10 & \dex{1.714}{4} & 38990 & 4.523 &
\dex{4.81}{46} & \dex{1.34}{46} & \dex{2.12}{45} \\
\dex{1.795}{7} & 2.84 & 10 & \dex{1.750}{4} & 39350 & 4.530 &
\dex{5.06}{46} & \dex{1.44}{46} & \dex{2.28}{45} \\
\dex{1.796}{7} & 2.82 & 10 & \dex{1.791}{4} & 39720 & 4.536 &
\dex{5.33}{46} & \dex{1.56}{46} & \dex{2.51}{45} \\
\dex{1.796}{7} & 2.80 & 10 & \dex{1.832}{4} & 40090 & 4.542 &
\dex{5.60}{46} & \dex{1.67}{46} & \dex{2.68}{45} \\
\dex{1.796}{7} & 2.79 & 10 & \dex{1.879}{4} & 40460 & 4.547 &
\dex{5.89}{46} & \dex{1.79}{46} & \dex{2.88}{45} \\
\dex{1.804}{7} & 2.74 & 10 & \dex{1.950}{4} & 41210 & 4.563 &
\dex{6.41}{46} & \dex{2.02}{46} & \dex{3.27}{45} \\
\dex{1.810}{7} & 2.68 & 10 & \dex{1.932}{4} & 41590 & 4.583 &
\dex{6.49}{46} & \dex{2.07}{46} & \dex{3.36}{45} \\
\dex{1.825}{7} & 2.61 & 10 & \dex{1.928}{4} & 42070 & 4.604 &
\dex{6.64}{46} & \dex{2.16}{46} & \dex{3.51}{45} \\
\dex{1.829}{7} & 2.61 & 10 & \dex{1.928}{4} & 42070 & 4.604 &
\dex{6.64}{46} & \dex{2.16}{46} & \dex{3.51}{45} \\
\dex{1.862}{7} & 2.68 & 10 & \dex{1.963}{4} & 41690 & 4.580 &
\dex{6.63}{46} & \dex{2.13}{46} & \dex{3.46}{45} \\
\dex{1.878}{7} & 2.75 & 10 & \dex{1.986}{4} & 41300 & 4.559 &
\dex{6.58}{46} & \dex{2.07}{46} & \dex{3.35}{45} \\
\dex{1.888}{7} & 2.83 & 10 & \dex{2.009}{4} & 40830 & 4.534 &
\dex{6.47}{46} & \dex{2.00}{46} & \dex{3.21}{45} \\
\dex{1.899}{7} & 2.91 & 10 & \dex{2.042}{4} & 40460 & 4.511 &
\dex{6.56}{46} & \dex{2.01}{46} & \dex{3.20}{45} \\
\dex{1.909}{7} & 2.98 & 10 & \dex{2.075}{4} & 40090 & 4.488 &
\dex{6.55}{46} & \dex{1.99}{46} & \dex{3.17}{45} \\
\dex{1.919}{7} & 3.06 & 10 & \dex{2.109}{4} & 39720 & 4.465 &
\dex{6.50}{46} & \dex{1.94}{46} & \dex{3.07}{45} \\
\dex{1.927}{7} & 3.16 & 10 & \dex{2.143}{4} & 39260 & 4.438 &
\dex{6.40}{46} & \dex{1.87}{46} & \dex{2.93}{45} \\
\dex{1.938}{7} & 3.93 & 10 & \dex{2.000}{4} & 34590 & 4.248 &
\dex{3.25}{46} & \dex{1.43}{45} & \dex{9.46}{43} \\
\dex{1.940}{7} & 5.25 & 10 & \dex{2.104}{4} & 30340 & 3.998 &
\dex{8.90}{45} & \dex{1.08}{43} & \dex{8.24}{39} \\
\dex{1.943}{7} & 11.4 & 10 & \dex{1.946}{4} & 20180 & 3.324 &
\dex{8.52}{43} & \dex{1.04}{39} & \dex{3.32}{34} \\
\dex{1.943}{7} & 12.3 & 10 & \dex{1.928}{4} & 19360 & 3.256 &
\dex{7.27}{43} & \dex{9.12}{38} & \dex{6.10}{33} \\
\dex{1.943}{7} & 15.9 & 10 & \dex{1.897}{4} & 16980 & 3.035 &
\dex{2.04}{43} & \dex{1.75}{40} & \dex{4.64}{34} \\
\hline
\end{tabular}
\end{table*}

\end{document}